\documentclass[twocolumn, pra, aps, superscriptaddress,floatfix]{revtex4-1}
\pdfoutput=1
\usepackage[caption=false,subrefformat=parens,labelformat=parens]{subfig}
\usepackage{graphicx}
\usepackage{float}
\usepackage{siunitx}
\usepackage{amsmath}
\newcommand{\angstrom}{\mbox{\normalfont\AA}}
\usepackage{bm}
\usepackage{multirow}
\usepackage{xcolor}
\usepackage{siunitx}
\usepackage{xfrac}

\usepackage[colorlinks=true, allcolors=blue]{hyperref}
\usepackage{color}

\usepackage{chemformula}
\usepackage{siunitx}

\def\k{{\bf k}}

\def\PB5{ \ch{Be5Pt}}
\begin{document}

\title{Remarkable low-energy properties of the  pseudogapped semimetal\PB5}

\author{Laura Fanfarillo}
\affiliation{Department of Physics, University of Florida, Gainesville, Florida 32611, USA}
\affiliation{Scuola Internazionale Superiore di Studi Avanzati (SISSA), Via Bonomea 265, 34136 Trieste, Italy}
\author{J.\ J.\ Hamlin}
\affiliation{Department of Physics, University of Florida, Gainesville, Florida 32611, USA}
\author{R.\ G.\ Hennig}
\affiliation{Department of Materials Science and  Engineering, University of Florida, Gainesville, Florida 32611, USA}
\affiliation{Quantum Theory Project, University of Florida, Gainesville, Florida 32611, USA}
\author{Ajinkya C.\ Hire}
\affiliation{Department of Materials Science and  Engineering, University of Florida, Gainesville, Florida 32611, USA}
\affiliation{Quantum Theory Project, University of Florida, Gainesville, Florida 32611, USA}
\author{P.\ J.\ Hirschfeld}
\affiliation{Department of Physics, University of Florida, Gainesville, Florida 32611, USA}
\author{Jungsoo Kim}
\affiliation{Department of Physics, University of Florida, Gainesville, Florida 32611, USA}
\author{Jinhyuk Lim}
\affiliation{Department of Physics, University of Florida, Gainesville, Florida 32611, USA}
\author{Yundi Quan}
\affiliation{Department of Physics, University of Florida, Gainesville, Florida 32611, USA}
\affiliation{Department of Materials Science and  Engineering, University of Florida, Gainesville, Florida 32611, USA}
\affiliation{Quantum Theory Project, University of Florida, Gainesville, Florida 32611, USA}
\author{G.\ R.\ Stewart}
\affiliation{Department of Physics, University of Florida, Gainesville, Florida 32611, USA}
\author{Stephen R. Xie}
\affiliation{Department of Materials Science and  Engineering, University of Florida, Gainesville, Florida 32611, USA}
\affiliation{Quantum Theory Project, University of Florida, Gainesville, Florida 32611, USA}
\date{\today}
\begin{abstract}
 We report measurements and calculations on the properties of the intermetallic compound \ch{Be5Pt}.
 High-quality polycrystalline samples show a nearly constant temperature dependence of the electrical resistivity over a wide temperature range.
 On the other hand, relativistic  electronic structure calculations  indicate the existence of a narrow pseudogap in the density of states arising from accidental approximate Dirac cones extremely close to the Fermi level. A small true gap of order $\sim$ 3 meV is present at the Fermi level, yet the measured resistivity is nearly constant  from low to room temperature.
 We argue that this unexpected behavior can be understood by a cancellation of the energy dependence of density of states and relaxation time due to disorder, and discuss a model for electronic transport.
 With applied pressure, the resistivity becomes semiconducting, consistent with theoretical calculations that show that the band gap increases with applied pressure.
 We further discuss the role of Be inclusions in the samples.

\end{abstract}

\maketitle
\section{Introduction }

Intermetallic compounds involving Be have not been the subject of intensive study, in part because of the element's toxicity.
However, several recent measurements in the Be-Pt series have displayed remarkable properties.
\ch{Be21Pt5} is a complex metallic alloy that crystallizes in the cubic space group $F{\bar 4}3m$, where polyhedral 26-atom Pt-Be clusters decorate four Wycoff sites, resulting in a 416-atom conventional unit cell.
Unlike most complex metallic alloys, this material displays superconductivity below a critical temperature of $T_c=\SI{2.06}{K}$~\cite{amon_cluster_2018}.

\PB5, on the other hand, crystallizes in the same space group but into a much simpler structure with only 24 atoms per conventional unit cell.
The electronic structure is  usually reported as semiconducting, with a gap of $ \SI{190}{meV}$ deduced from the high-temperature resistivity.  This is consistent with the energy scale found within density functional theory (DFT) \cite{amon_interplay_2019}   corresponding to the excitation of electrons from the top of the valence band to a sharp peak in the conduction band.  However, there is a general depletion of the density of states over a wider range, decreasing down to the Fermi level, and it is difficult to judge from the calculations presented in Ref.~\onlinecite{amon_interplay_2019} whether a true gap exists at some lower energy scale.
Experimentally, the resistivity is metallic but with a nearly temperature-independent behavior over the entire measured temperature range from \SI{3}{K} - \SI{300}{K}.
Though this suggests that the system is a metal at the lowest energies, quantitative estimates are difficult, as the samples contains a filamentary secondary phase of Be metal.
The low-temperature electrical resistivity of Be  ($\SI{6.4E-9}{\micro\ohm\cm}$) can be substantially smaller than that of Cu ($\SI{3.4E-7}{\micro\ohm\cm}$)~\cite{fickett_electrical_1982}.
The existence or nonexistence of a true gap, and the origin of the extremely flat temperature dependence of the resistivity below \SI{100}{K}, remain puzzles.  

 In the pioneering work of  Ref. \onlinecite{amon_interplay_2019}, the intrinsic resistivity of \PB5 crystals, $T$-independent over a 200 K range,  was reported on  samples tens of microns in size, and electronic structure calculations on a large energy scale were presented.  Here we follow up these intriguing results with a closer examination of the low-energy band structure, an analysis of the unusual transport behavior, and studies of the pressure dependence of electronic properties up to 30 GPa.  
 
First, using an arc melting technique, we have grown polycrystalline samples  which we characterize by x-ray diffraction, resistivity, Hall effect, and specific heat measurements. We present basic sample characterization of our   polycrystals  in Sec. \ref{sec:materials}.

We also have performed DFT calculations for this system, focusing  in Sec. \ref{sec:bandstructure} on the low-energy band structure near the Fermi level.
The unusual structure in the DOS turns out to be due to light-mass valence and conduction bands, consisting of Pt $d$ states and Be $p$ states, that nearly touch the Fermi level and exhibit a tiny indirect band gap of order only \SI{3}{meV}.  This makes \PB5 the only binary, nontopological intermetallic system with calculated band gap significantly smaller than room temperature~\cite{Jain2013}.  Even more intriguingly 
the two bands near the Fermi level provide a quasilinear DOS until nearby heavier mass bands are reached; these determine the  band gap  previously stated in the literature.
Finally, at slightly higher energies we note the existence of doubly degenerate Weyl loops within \SI{100}{meV} of the Fermi level.
This suggests that electron doping with, \textit{e.g.} relatively small amounts of Au, could create a system with Weyl states at the Fermi level.

Next, we provide in Sec. \ref{sec:transport} a simple model to explain the low temperature transport, including the remarkable fact that the resistivity is nearly flat in temperature, whereas the Hall coefficient decreases rapidly with decreasing temperature.
The model relies on the anomalous scattering behavior of conduction electrons from static disorder in a pseudogap state, analogous to Dirac metals like graphene, although the quasi-Dirac point in this system is accidental in nature.  

Finally, in Sec. \ref{sec:pressure} we present calculations and measurements of \PB5 under pressure.
DFT calculations in the $F{\bar 4}3m$  structure predict that the tiny gap near the Fermi level further opens with pressure, with concomitant enhancement of resistivity and eventually true semiconducting behavior.
This is precisely what is observed in the measurements reported here.
No structural transitions are anticipated theoretically, nor does the electrical resistivity data suggest such transitions up to \SI{30}{GPa}.

\section{Materials characterization }
\label{sec:materials}
\begin{figure}[!ht]
  \includegraphics[width=0.99\columnwidth]{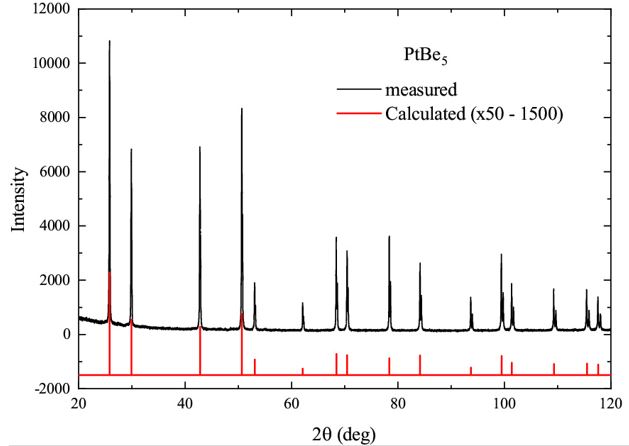}
  \caption{Measured and calculated x-ray pattern for  \ch{Be5Pt}.  The smaller peak on the high angle side of the measured higher angle lines is due to the incident Cu K-alpha xray radiation having two wavelengths (K-alpha1 and K-alpha2) while the calculated pattern in red is for a single incident wavelength.  A small offset has been introduced to displace the patterns.}
  \label{fig:xray}
\end{figure}

We prepared polycrystalline  samples of  \ch{Be5Pt} following a strategy guided by the phase diagram of Pt-Be, which shows the phases (starting from the Be-rich side of the phase diagram)  Be, \ch{Be5Pt}, and then possibly Be$_{21}$Pt$_5$ (or Be$_{4.2}$Pt)~\cite{amon_cluster_2018}.
The Be$_{21}$Pt$_5$ phase is described~\cite{amon_cluster_2018} as superconducting (resistive onset) at $T_c = \SI{2.06}{K}$.
When we prepared \ch{Be5Pt} via arc-melting, we therefore added extra Be (which has a high vapor pressure  at the elevated melting temperature) to compensate for the Be mass loss during the melting process.
As long as the sample remains Be-rich after arc-melting, the x-ray characterization shows  single phase cubic \ch{Be5Pt} structure  (Fig.~\ref{fig:xray}), { without detectable impurity phases}

\begin{figure}[!ht]
  \includegraphics[width=0.69\columnwidth]{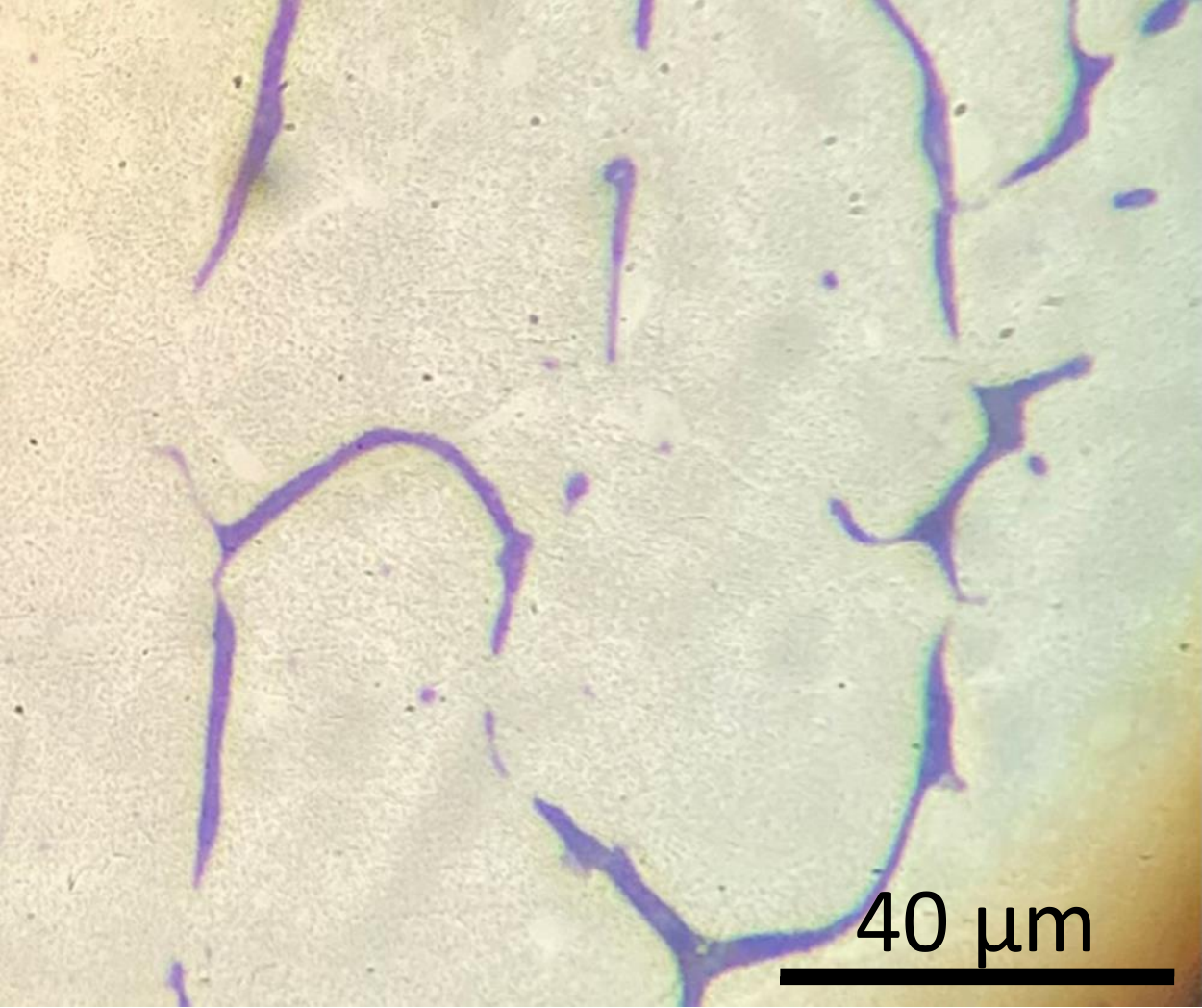}
  \caption{600x magnification image of a slightly Be-rich ($\sim 5\%$) \ch{Be5Pt} sample showing Be second phase regions (blue/purple).}
  \label{fig:micro}
\end{figure}

\begin{figure}[!ht]
  \includegraphics[width=0.99\columnwidth]{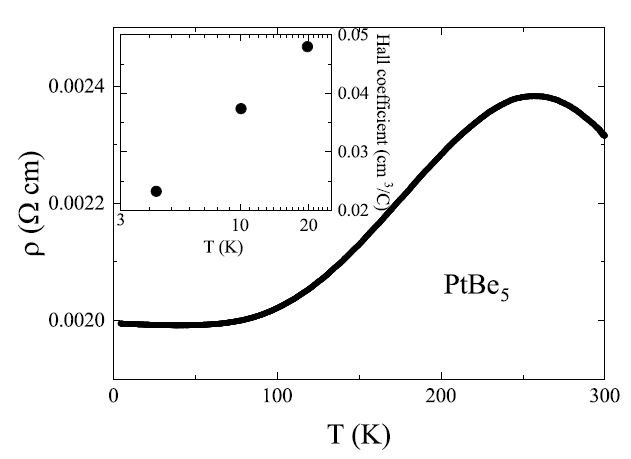}
  \caption{Resistivity and low temperature Hall coefficient (insert). }
  \label{fig:transport_expt}
\end{figure}

\vskip .2cm
If sufficient Be is evaporated in the production process to produce a sample more Be poor than the 5:1 ratio in \ch{Be5Pt}, the x-ray pattern becomes much more complicated, and superconductivity (absent in the stoichiometric or super-stoichiometric \ch{Be5Pt} samples) with a $T_c$ resistive onset of \SI{2.47}{K} is detected.
We presume this is indicative of the presence of a minor amount of the Amon, \textit{et al.} Be$_{21}$Pt$_5$ phase~\cite{amon_cluster_2018}, although our $T_c$ onset is slightly higher.
In our typical preparations of Be-rich \ch{Be5Pt}, the excess Be tends to form connected regions (see micrograph in Fig.~\ref{fig:micro}), leading to a low temperature electrical resistivity significantly lower than ‘ideal’ stoichiometric \ch{Be5Pt}, where according to Amon \textit{et al.}~\cite{amon_interplay_2019}, $\rho_0$ as $T\rightarrow 0$ is $\sim$~360 m$\Omega$-cm.  In Fig. \ref{fig:transport_expt}, we show the resisitivity of a sample of \PB5, together with the low-temperature Hall coefficient as a function of temperatures.   Our samples of Be-rich \ch{Be5Pt} have $\rho_0$ as $T\rightarrow 0$ between 0.45 and 2.0 m$\Omega$-cm.

The presence of 5 percent second phase of Be threaded through the sample as illustrated in Fig. \ref{fig:micro} is not an impediment to determining the temperature dependence of the majority phase - see Fig. \ref{fig:transport_expt}. Similarly, the temperature dependence of the Hall effect (inset to Fig. \ref{fig:transport_expt}) and the pressure response of the resistivity (discussed below in section VI) can also be determined.  The essential point is that the resistivity of Be\cite{Be_rho_1,Be_rho_2} is so much smaller than the intrinsic resistance of \PB5\cite{amon_interplay_2019} that the measured resistance of our composite sample, treated in a resistor network model, would be much smaller than our measured value of 2 m$\Omega$-cm if the Be percolated across the sample.


\vskip .5cm

\section{Electronic structure }
\label{sec:bandstructure}
\subsection{Structure and Method}
\begin{figure}[!ht]
  \includegraphics[width=0.58\columnwidth]{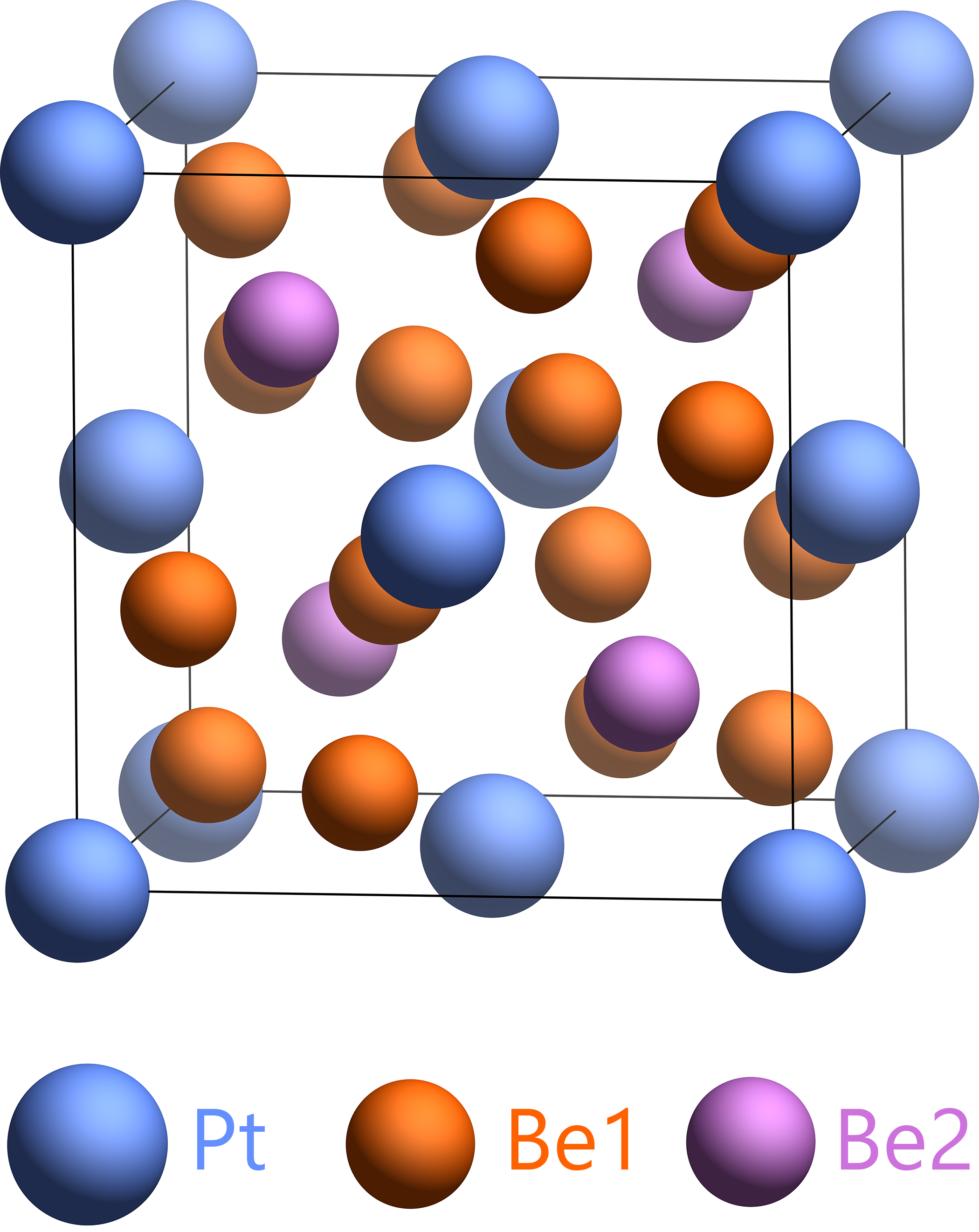}
  \caption{The crystal structure of \ch{Be5Pt} is face-centered cubic. Be atoms sit at two non-equivalent Wyckoff positions 16e (Be1) and 4c (Be2). Be1 form networks of corner sharing tetrahedra.\cite{amon_interplay_2019} The Be1-Be1 bond length is 2.10 \si{\angstrom}, while Be1-Be2 bond length is 2.48 $\si{\angstrom}$. The distance between two nearest Be2 atoms is 4.22 \si{\angstrom}. }
\label{str} 
\end{figure}
\ch{Be5Pt} crystallizes in the non-centrosymmetric space group $F\bar{4}3m$ ($\#216$). There are four formula units per conventional cell. Pt, Be1 and Be2 sit at Wyckoff positions 4a, 4c and 16e respectively. See Fig.~\ref{str}. The linearized augmented plane wave method, as implemented in WIEN2k\cite{WIEN2k}, is used to carry out first-principles calculations. The lattice constant of \ch{Be5Pt} is \SI{5.975}{\angstrom}.  The generalized gradient approximation (GGA) \cite{PhysRevLett.77.3865} is chosen as the exchange correlation functional.  Relativistic effects are included and spin-orbit coupling is treated with the second variational method.
Muffin-tin radii of Pt and Be are 2.47 and 1.98 a.u., respectively. R$_{MT}^{min}$K$_{max}$, which determines the plane wave cutoff in the interstitial region, is kept at 7. To properly describe the electronic structure of \ch{Be5Pt} near the Fermi level, we adopt a $\Gamma$-centered k-mesh with 11921 k-points in the irreducible Brillouin zone wedge. For the structure prediction calculations, we used the Vienna Ab initio Simulation Package (VASP). The electronic structures and the band gaps of Be$_5$Pt obtained using VASP and WIEN2k are similar.
\begin{table}
\begin{tabular}{cccccc}
\hline
  & Site & x & y & z \\
\hline
\hline
Pt & 4a & 0. & 0. & 0. \\
Be1 & 16e & 0.124 & 0.124 & 0.624\\
Be2 & 4c & 0.25 & 0.25 & 0.25\\
\hline
\end{tabular}
\caption{Coordinates of Pt and Be under ambient pressure obtained from DFT structure relaxation. }
\end{table}

\subsection{Results and discussions}
\begin{figure}[!ht]
  \includegraphics[width=0.99\columnwidth]{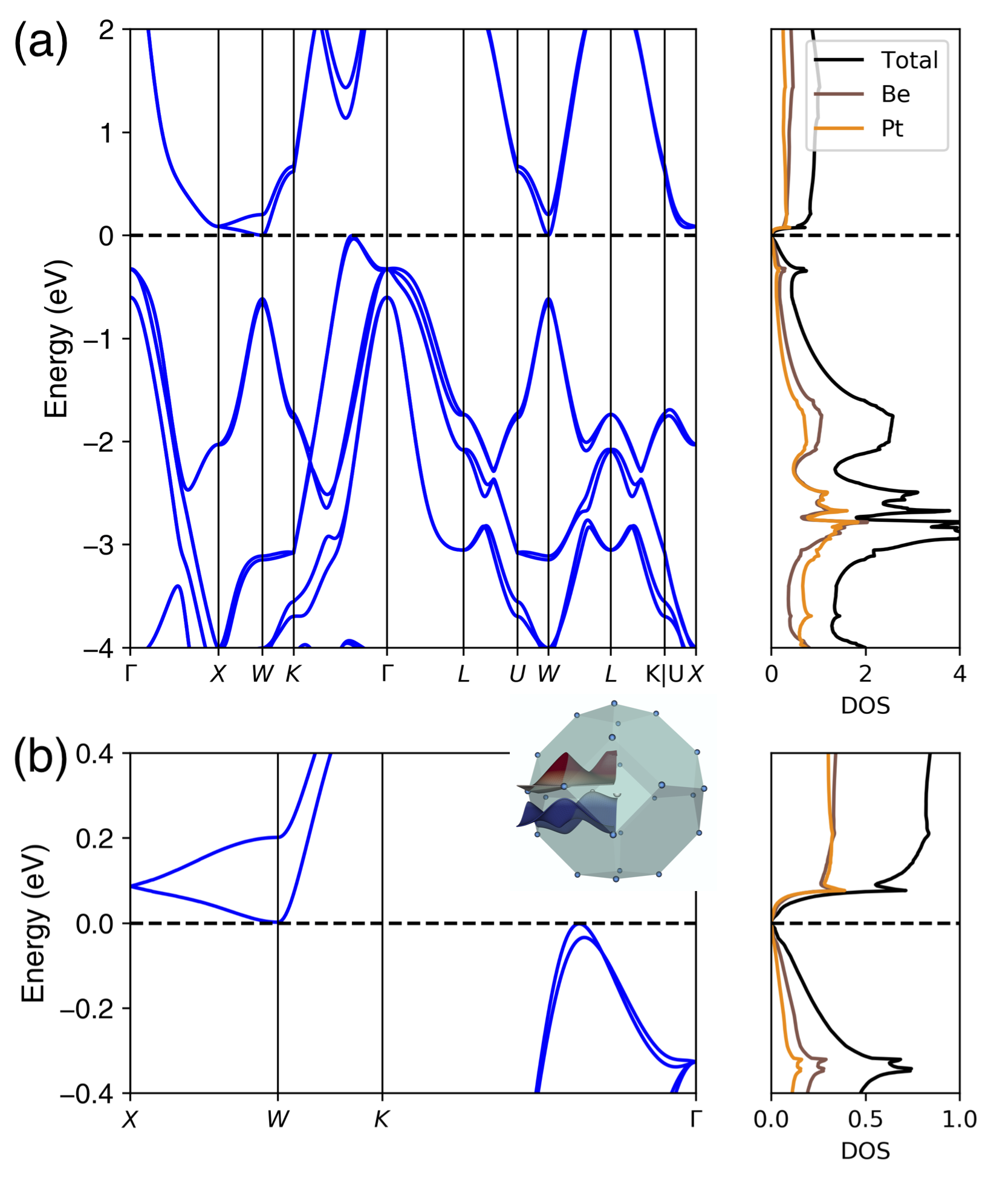}
  \caption{(a) electronic structure (left) and density of states (right) of \ch{Be5Pt} from DFT. (b) electronic structure (left) and density of states of \ch{Be5Pt} within 0.4 eV of the Fermi level. DOSs are in units of states per eV per primitive cell. The two lowest conduction bands are linear along $X\rightarrow W$ and they touch at the X point (Weyl point).}
  \label{fig:es}
\end{figure}
The calculated density of states (DOS) of \ch{Be5Pt} has a ``pseudogap" $E_{pg}$ of roughly 400meV, where the DOS is strongly suppressed, and in addition, a true  gap $\Delta$ of $\sim 3$ meV (see Fig.\ref{fig:es}).
Low temperature transport properties of \ch{Be5Pt}  are therefore
dominated by the states in close vicinity to the Fermi level, 
\textit{i.e.} the states near the valence band maximum and the 
conduction band minimum. The valence band maximum at (0, 0.28, 0.28) and
its equivalent k-points are trivial $M_3$ type van Hove singularities 
with light effective masses $m_x=0.27 m_e$, $m_y=0.15m_e$ and $m_z=0.17m_e$. 
At the $W$ point, the lowest two unoccupied states are split by \SI{200}{meV} due to spin-orbit coupling.
In addition, dispersion of the lowest unoccupied conduction band near $W$ is strongly anisotropic, with relatively flat dispersion along $W\rightarrow X$ and a steep slope along $W\rightarrow K$. 

\begin{figure}[!h]
  \centering
  \includegraphics[width=0.99\columnwidth]{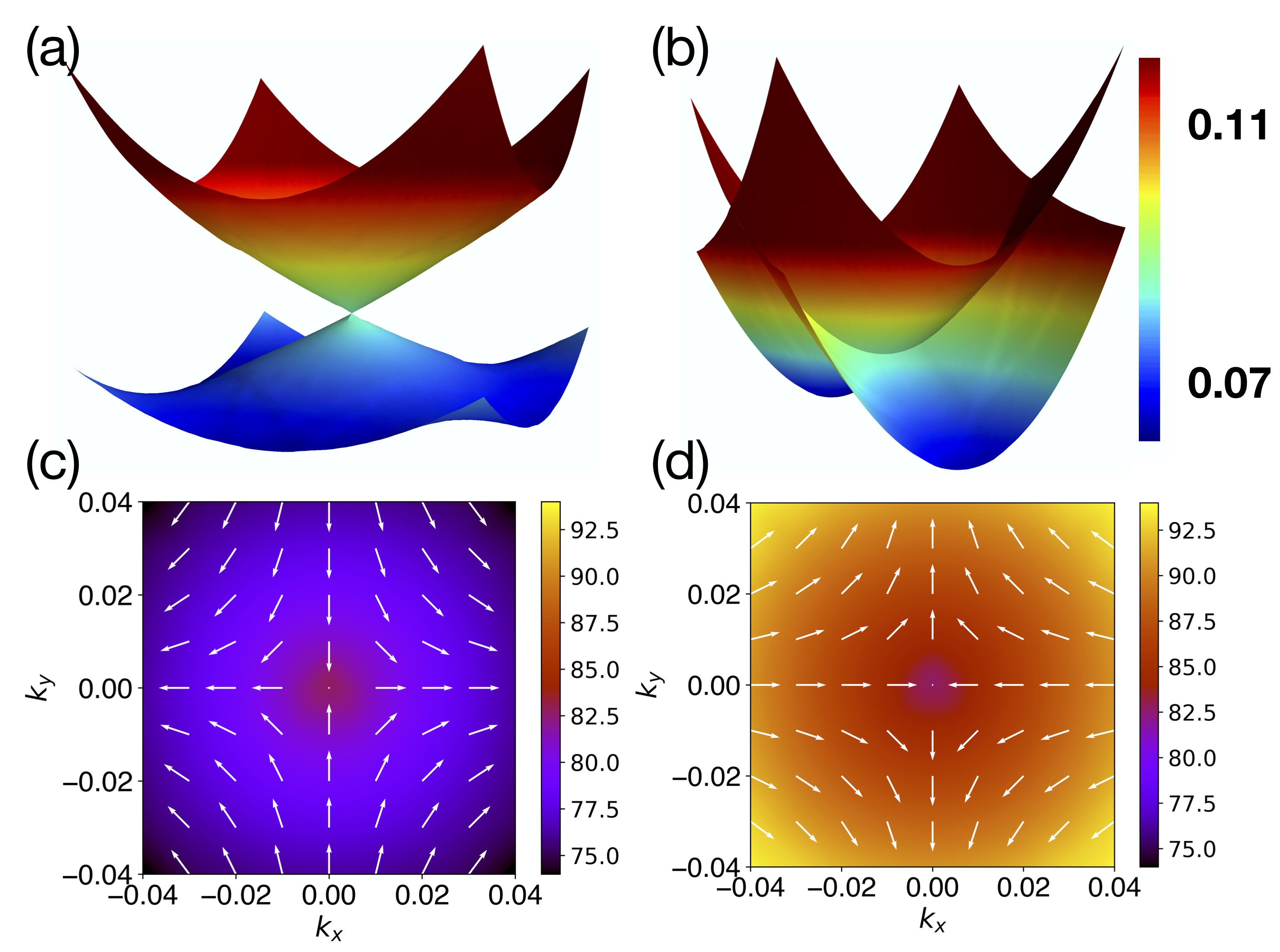}
  \caption{(a) DFT low energy dispersion of the two lowest unoccupied bands  on the $k_x-k_y$ plane, i.e. the plane that's perpendicular to the $\Gamma-X$ direction.  (b) band dispersion of the two lowest unoccupied bands on the $k_x$-$k_z$ plane. Dresselhaus spin-orbit coupling vanishes along $\Gamma-X$ which leaves the two fold spin degeneracy intact. Spin textures of the lowest and the next lowest unoccupied bands are shown in (c) and (d).   } 
  \label{fig:disx}
\end{figure}

\ch{Be5Pt} crystallizes in space group ($\#216$) with broken inversion symmetry. 
The inversion symmetry breaking of the bulk \ch{Be5Pt} results in Dresselhaus spin-orbit
coupling (SOC) which lifts spin degeneracy at general k-points in the Brillouin 
zone. Symmetry analysis of space group ($\#216$) \cite{PhysRev.100.580} reveals 
that SOC vanishes along $\Gamma\rightarrow X$ path, which our DFT calculations
confirm. Bands along $\Gamma\rightarrow X$ are therefore doubly degenerate, see Fig.~\ref{fig:disx} (b). In Fig.~\ref{fig:disx} (a), the band dispersions of the two lowest
conduction bands near $X$ are shown, demonstrating the linear dispersion near the Weyl point at $X$.

Because the DOS has a sharp edge above the Fermi level, a small amount of electron doping can shift states near $W$ and $X$ below the Fermi level. In Fig.~\ref{fig:disx}, we plot the band dispersions of the two lowest conduction bands
near the $X$ point on the $k_x-k_y$ plane. We have chosen (001) as the $X$ point. Interestingly, local band dispersions on $k_x-k_y$ plane are linear near $X$, which is in agreement with
Ref. \cite{PhysRevB.72.195347}.  The effective Hamiltonian near $X$ can therefore be written as
\begin{eqnarray}
H_{D} = \beta( k_x \sigma_x - k_y \sigma_y)
\end{eqnarray}
$\sigma_x$ and $\sigma_y$ are Pauli matrices and $k_{x/y}$ are crystal momenta along conventional reciprocal lattices. As a result of the linear Dresselhaus SOC, the expectation values of $s_x$ and $s_y$ are dependent on $k_x$ and $k_y$. We compute the spin textures of the two lowest conduction bands using the WIEN2k code\cite{PhysRevB.94.125134} and find that the DFT spin textures are exactly what would be obtained from the linearized $H_D$ in Eq. (1).

To understand the effect of pressure on the electronic structure of \ch{Be5Pt}, we carry out DFT calculations of \ch{Be5Pt} under \SI{5}, \SI{10}, \SI{20}, \SI{30}, \SI{40}, \SI{50}, \SI{70}, \SI{90}, \SI{110}, \SI{130} and \SI{150} {GPa}. Under ambient pressure, the indirect band gap of \ch{Be5Pt} is only 3 meV. As pressure increases from 0 to 70 GPa, the indirect band gap goes up quickly to 78 meV. Above 70 GPa, it starts to slowly decrease with higher pressure. The pseudogap, on the other hand, increases with pressure monotonically from 0.4 eV under ambient pressure to 1.2 eV under 150 GPa. 

We note that pressure has a dramatically different impact on occupied vs unoccupied states near the Fermi level. The occupied states within 0.25 eV of the Fermi level are insensitive to changes in pressure. For the unoccupied states within 1 eV above the Fermi level, pressure has the effect of transferring a significant amount of spectral weight away from the Fermi level, thereby widening the pseudogap.

 It is interesting to consider what may happen to the properties of this material with a small amount of electron doping, \textit{e.g.} with Au.  First, we expect the electronic properties to change rather rapidly, due to the steep rise in the calculated DFT density of states on the electron doped side at ambient pressure.  These calculations suggest that as little as 2\% Au could change the density of states at the Fermi level by orders of magnitude.  In addition, slightly more Au doping should raise the chemical potential to the $\sim \SI{0.1}{eV}$ level where the Weyl point would be at the Fermi surface, assuming a rigid band shift.  We plan to study this doping sequence in a subsequent study.
\begin{figure}[!h]
  \includegraphics[width=\columnwidth]{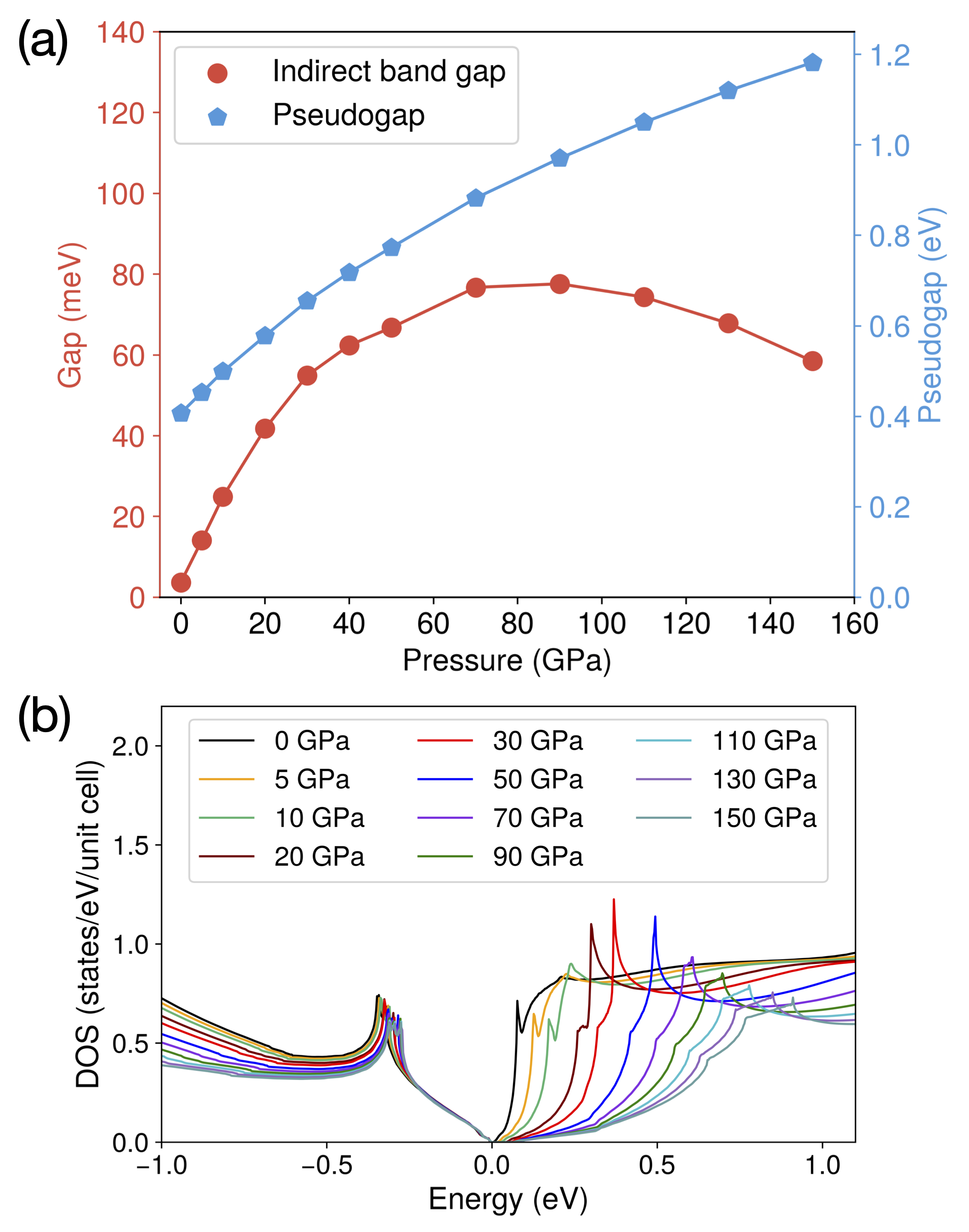}
  \caption{DFT calculations of (a) the indirect band gap (red) and the pseudogap (blue) of \ch{Be5Pt} versus pressure. 
  (b) DOS of \ch{Be5Pt} under pressure.  }
  \label{fig:esp}
\end{figure}

\section{Interpretation of transport }
\label{sec:transport}

The rather high value of the measured resistivity of our samples is consistent with a semiconducting behavior, as claimed in Ref. \cite{amon_interplay_2019}, but the temperature dependence of the resistivity, constant over more than \SI{100}{K} at low $T$, is not.  Electronic structure calculations presented in Sec. III (see Fig. \ref{fig:es}) suggest, in fact, an asymmetric pseudogap in the density of states from about \SI{-0.35}{eV} to \SI{0.05}{eV}, with a peak in the unoccupied density of states at the upper edge created by a flat band.  Closer to the Fermi energy, a tiny true indirect gap of about 3 meV opens.  
At first glance, such a pseudogap/full-gap combination might be expected to  lead to  a semiconducting temperature dependence of  the DC transport, but we argue here that this may be misleading.  

\begin{figure}[!h]
  \includegraphics[width=\columnwidth]{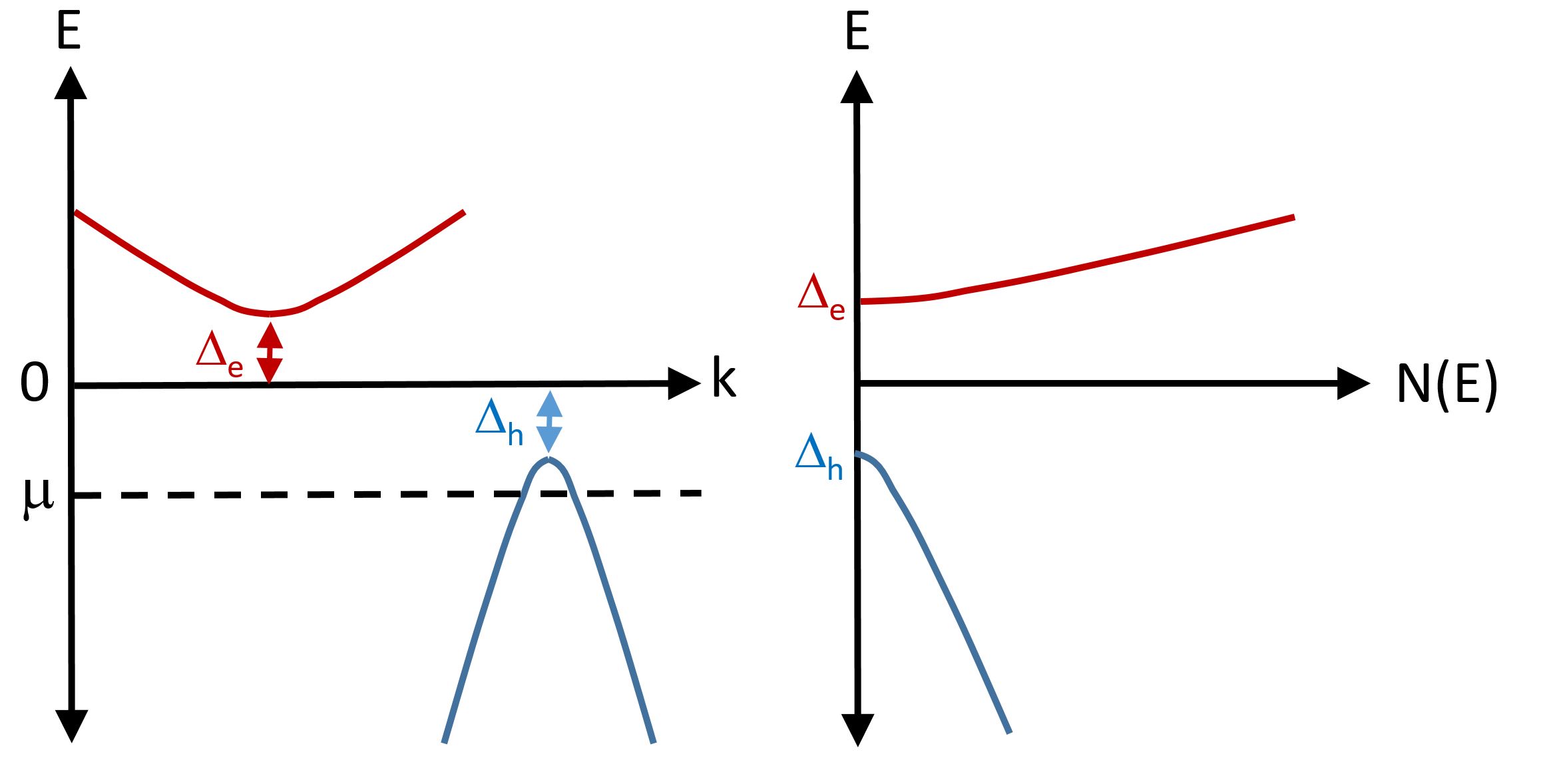}
  \caption{Cartoon of \PB5 band structure near Fermi level at ambient pressure. }
  \label{fig:schematic_bands}
\end{figure}

Let us assume that we can model the low energy physics by an asymmetric, very slightly gapped V-shaped density of states with a tiny true gap, $\Delta=\Delta_e+\Delta_h \sim \SI{3}{meV}$, as illustrated in the schematic Figure \ref{fig:schematic_bands}.  
We model the electronic structure relevant for low-temperature transport with two Dirac-like bands: an electron-like band, $\epsilon^{e}_\k = \gamma_{e}  k$ and a hole-like one, $\epsilon^{h}_\k = - \gamma_{h}  k$,
where $\k$ is measured with respect to the near-touching point of the band and we assume $\gamma^h > \gamma^e$. Such a model yields to a density of state linear in energy $N(\epsilon_\k)\sim \epsilon_\k$ 
similar to the band structure calculations. Comparing with Fig.~\ref{fig:schematic_bands}, the model discussed in this section corresponds to zero gap $\Delta_\alpha = 0$, $\alpha=e,h$, and zero chemical potential $\mu$, but it can be easily generalized  to include hole doping or gapping the band, both of which are discussed in the Appendix. 

Electronic transport in a typical metallic system with large nonzero density of states at the Fermi level may normally be well-described by assuming a constant carrier scattering time $\tau$; this leads to a temperature-independent resistivity and Hall coefficient.  Both  the band structure calculations and the experimental Hall data for this system presented here suggest that this picture is incorrect.  In fact, since the density of states is strongly energy-dependent near the Fermi level,  a constant relaxation time approximation is no longer valid.  Let us assume that the  transport at low temperatures is dominated by a small number of weakly scattering point-like defects with concentration $n_i$ and potential $u$. As shown in the Appendix the scattering rate from pointlike weak impurities for such a model is given by~\cite{Ando98}
 \begin{eqnarray}\label{Eq:selfenergy0}
 {1\over \tau_{\alpha}(\omega)}&=& 2\Sigma''(\omega) = \Gamma^0_{\alpha} |\omega|,
 \end{eqnarray}
where we introduced the dimensionless parameter $\Gamma^0_\alpha = n_i u^2/\gamma^2_{\alpha}$, characterizing the scattering strength. Note that we have set both the lattice parameter $a=1$ and the reduced Planck constant $\hbar=1$ in this and subsequent expressions. 

The longitudinal conductivity is given by the sum over the bands of the hole and electron contributions that are given by
\begin{eqnarray} \label{sigma_xx_v2}
\sigma^{\alpha}_{xx} &\sim& \frac{e^2 }{2  \pi \Gamma^0_\alpha},
\label{eq:sxx}
\end{eqnarray}
%
From (\ref{eq:sxx}), we see that the $T$-dependence of the $H=0$ longitudinal conductivity vanishes to leading order, as expected. Note there is no universal term in the conductivity, due to the neglect of self-consistency in the Born approximation, which is relevant only over an exponentially small temperature range.
The $T$-dependence of the conductivity  from inelastic scattering will of course contribute at sufficiently high temperatures, but this depends on details of the system, such as Debye frequency, phonon distribution, etc. that we do not model here.  

The analysis of the Hall coefficient requires the computation of the transverse component of the conductivity tensor, $\sigma_{xy}$. Also in this case, the total conductivity is given by the sum of the electron and hole contributions. As detailed in the Appendix and discussed in \cite{Ando2006}, the cyclotronic frequency $\omega_c$ for a Dirac-like electronic structure is singular at zero energy, thus the denominator $(1+ \omega^2_c \tau^2)$ cannot be neglected even if the magnetic field is small. Details of the derivation can be found in the Appendix, the final result for the transverse conductivity of the $\alpha$-band reads
\begin{eqnarray}
\sigma^\alpha_{xy} &=& \pm \frac{e^2}{2 \pi \Gamma^0_{\alpha}} \ \kappa_{\alpha}^2 \int_{-\infty}^{0} d\epsilon \ \frac{\epsilon^2}{ \epsilon^4  + \kappa_{\alpha}^4 } \  \bigg(-\frac{\partial f}{\partial \epsilon }\bigg).
\end{eqnarray}
where the parameter $\kappa_\alpha$ has the dimensions of  energy and it is defined as $\kappa_\alpha^2 = e H v_\alpha^2 / \Gamma^0_\alpha$ with $v_\alpha =\gamma_\alpha$ the band velocity.
The denominator determines the $\sigma_{xy}\sim T^2 $ dependence as $T\rightarrow 0$ and 1/$T^2$ as $T\rightarrow \infty$. The crossover between low and high $T$ behaviors depends on the value of $\kappa_\alpha$.
Now we discuss combining the contributions of hole and electron bands. Notice that in a completely symmetric model for the Dirac bands, the system would be compensated i.e. $\sigma_{xy}^h = -\sigma^e_{xy}$ and the Hall coefficient would be zero. However, in our model in order to reproduce the particle-hole asymmetry of the density of states we  assume $\gamma^h>\gamma^e$, as a consequence here $\kappa^h>\kappa^e$.  The asymmetry of the Dirac bands of our model implies that the total conductivity is  dominated by the hole contribution since $\gamma_h > \gamma_e$.  This effect is enhanced by the $T$-dependence of the chemical potential, which  moves to negative values as $T$ increases (see Appendix).

Despite the fact that $\sigma_{xy}$ increases with increasing $T$, a consistent explanation of the Hall coefficient temperature dependence reported in Fig. \ref{fig:transport_expt}  is difficult using this model alone. This is because the Hall coefficient is given by 
\begin{equation}
R_H={\sigma_{xy}\over H(\sigma_{xx}^2+\sigma_{xy}^2)},
\label{eq:RH}
\end{equation}
and the $T$ dependence of $R_H$ depends not only on $\sigma_{xy}$, but in principle also on $\sigma_{xx}(H)$ (see Eq.\ref{app:sigmaxx_h}). Applying Eq. (\ref{eq:RH}) directly with the current model  dominated by the hole band, 
 gives an $R_H$ with positive sign that increases as $T\rightarrow 0$ in contrast to the experiment. The strong $T$ dependence of $\sigma^h_{xx}(H)\sim T^4$ at low $T$ is at the root of this discrepancy. As discussed in the Appendix, the inclusion of a small gap in the model does not changes the Hall coefficient $T$-behavior. However, if one assumes the Fermi level lies a few meV inside the electron band, we recover a non-negligible contribution of the electron band at low temperature. This effectively cuts off the singular behavior of $R_H$ at low temperature and, within a certain range of parameters, and can reproduce a $R_H(T)$ similar to the one experimentally observed as shown the Appendix.
 
However this result requires fine tuning, which is probably unreasonable given the other uncertainties in the analysis. The main source is the existence of Be inclusions in the samples, as shown in Fig. \ref{fig:micro}. 
Although the precise values of the resistivity of the Be inclusions in our samples are unknown, they clearly correspond to a conductivity much larger~\cite{Be_rho_1,Be_rho_2} than the intrinsic \ch{Be5Pt} material, in agreement with the conclusions of Ref. ~\onlinecite{amon_interplay_2019}.  If we crudely model our samples as consisting of two parallel conducting channels, we expect that $\sigma=\sigma^{\rm Be}+\sigma^{\rm \PB5}$, with $\sigma^{\rm Be}>>\sigma^{\rm \PB5}$.  Note $\sigma^{\rm Be}$ is the effective conductivity of the Be conducting network.  Because $\sigma^{\rm \PB5}_{xx}$ is also $T$-independent and smaller than $\sigma_{xx}^{\rm Be}$, we may neglect it entirely in the estimation of $R_H$, and assume   $\sigma_{xx}\approx \sigma_{xx}^{\rm Be}$,  quite $T$-independent over 100K, and furthermore only weakly dependent on $H$ due to its strong metallic character.  On the other hand, if $\sigma_{xy}^{\rm Be}$ is not too much larger than $\sigma_{xy}^{\rm \PB5}$, its temperature dependence will dominate the extremely weak temperature dependence of $\sigma_{xy}^{\rm Be}$.  Furthermore, while the Hall coefficient of pure hcp Be metal has a positive or negative sign according to the direction of the magnetic field perpendicular to or parallel to the basal plane\cite{Be_Hall}, the former value is signifiantly larger, and is expected to dominate in the Be inclusions present here; we therefore assume that $\sigma_{xy}^{\rm Be}$ is constant in $T$ and has a positive sign, like $\sigma_{xy}^{\rm \PB5}$, which is dominated by the light hole states. The temperature dependence of the latter dominates, however, so we conclude that for our samples, 
\begin{equation}
R_H\approx {\sigma_{xy}^{\rm Be} +\sigma_{xy}^{\rm \PB5}\over H ({\sigma_{xx}^{\rm Be}})^2} \sim 1 + c T^2,
\end{equation}
where $c$ is a positive constant.  This result agrees  qualitatively with the measured Hall coefficient of the composite sample.  If samples can eventually be prepared without the Be inclusions, the Appendix contains more detailed predictions for the expected intrinsic behavior.
\vskip 1cm

\section{  Calculations under pressure }

We used the Genetic Algorithm for Structure and Phase Prediction (GASP)\cite{Tipton_2013,revard2016grand} to search for possible phase transitions in \ch{Be5Pt} under pressure at fixed composition. Two GASP  searches were performed at 50 GPa and 150 GPa. The ambient pressure atomic structure, along with 20 randomly generated structures, were used to  initialize the GASP searches. GASP uses a genetic algorithm to perform global optimization, minimizing the enthalpy of formation in candidate structures. The algorithm iteratively creates ‘offspring’ structures from ‘parent’ structures by using genetic operators like mutation  or mating, corresponding to operations like adding and removing atoms or splicing two structures, respectively. Structural features or ‘genes’ which tend to lower the enthalpy of a structure are promoted  in later structures as the algorithm progresses.

VASP\cite{KRESSE199615}\cite{PhysRevB.54.11169} was used to relax the structures and calculate the  enthalpies. The cutoff energy for  the plane-wave basis set was set to 520 eV. {We used the projector augmented wave (PAW) pseudopotentials\cite{PhysRevB.50.17953} and the Perdew-Burke-Ernzerhof (PBE) generalized gradient approximation\cite{PhysRevLett.77.3865} (GGA) for the exchange-correlation functional.} A k-point density of 20 k-points per 	\si{\angstrom}\textsuperscript{-1}  was used for all the DFT relaxations. {A stopping criterion of 600 DFT relaxations was used for both GASP  searches. Setting the maximum number of atoms in generated structures limits the search space to structures containing 6 or 12 atoms. In a previous work by the authors of GASP, it was noted that increasing the maximum system size also exponentially increases the number of local minima in the energy landscape \cite{Revard2014}. Efficiency is further reduced because individual energy calculations are much more expensive for larger structures. In another work by  Lyakhov et al., it was noted that randomly-generated large cells often have very poor formation energies; many are glass-like structures \cite{Lyakhov2010}. Based on these points, we have enforced a maximum of 12 atoms per generated configuration, equivalent to two formula units, based on available computational resources.}

\begin{figure}
    \centering
    \includegraphics[width=\columnwidth]{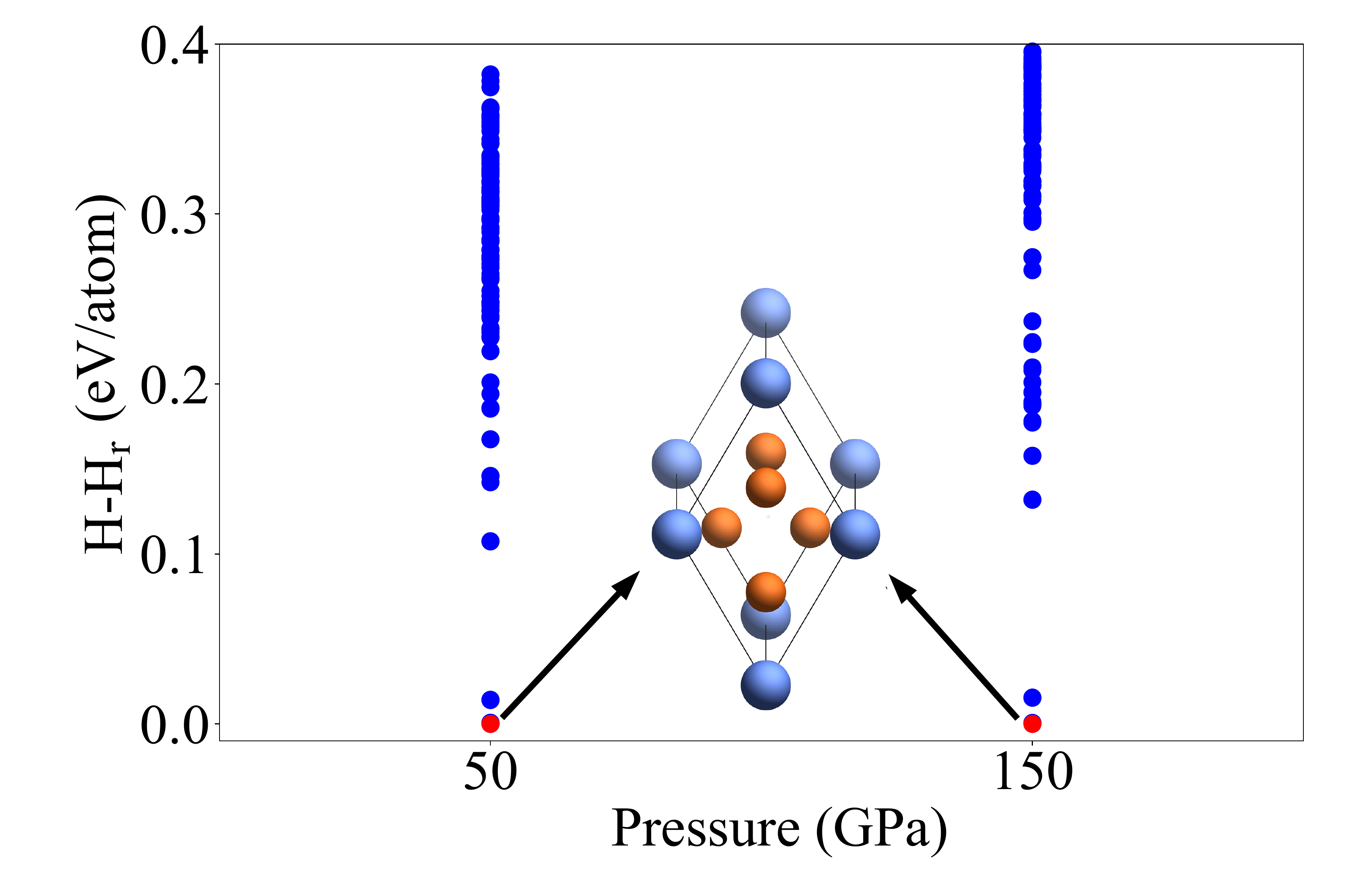}
    \caption{Enthalpy difference of structures generated in GASP at 50 GPa and 150 GPa with respect to the lowest enthalpy structure, H$_r$, shown in red. All of the low enthalpy structures have $F\Bar{4}3m$ spacegroup.  The inset shows the primitive lattice with the lowest enthalpy.}
    \label{fig:gasp_fig}
\end{figure}

\begin{figure*}
    \centering
    \includegraphics[width=\textwidth]{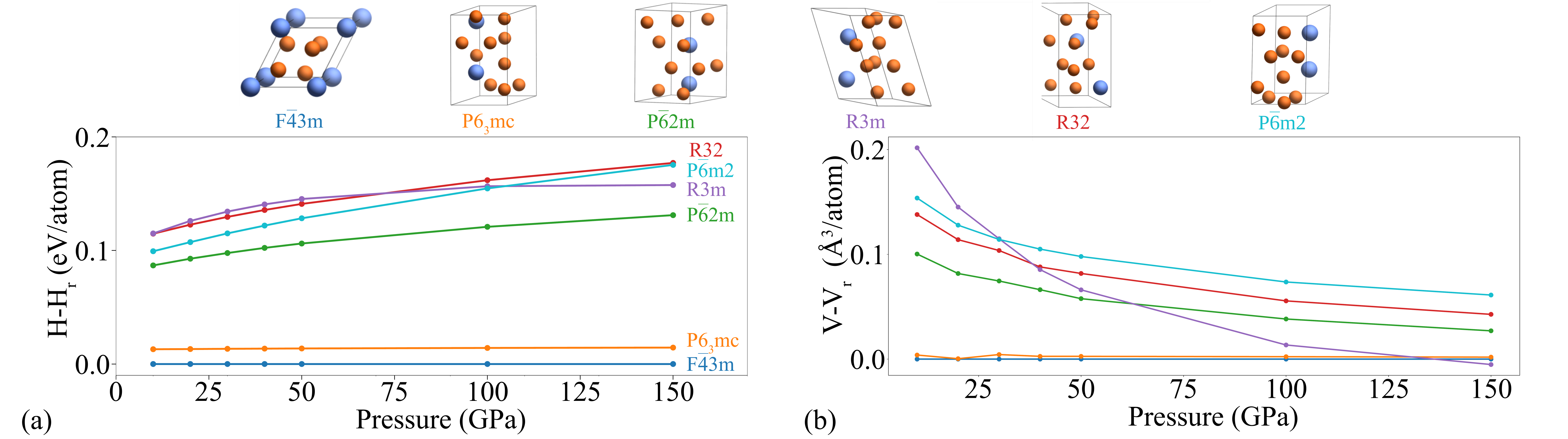}    \caption{(a) Enthalpy  and (b) volume relative to the $F\bar 4 3m$ ambient condition ground state structure as a function of pressure of the lowest enthalpy structures from the GASP structure search at \SI{50}{GPa} and \SI{150}{GPa}. }
    \label{fig:press_study}
\end{figure*}

Fig.~\ref{fig:gasp_fig} shows the enthalpy of structures generated by GASP  at 50 GPa and 150 GPa. During the GASP searches, structures that were predicted to be more stable relaxed to structures with $F\Bar{4}3m$ spacegroup, which is the same as that of the ambient condition ground-state structure.  This indicates that no structural phase transition is seen in PtBe5 under pressure at 50 GPa and 150 GPa. Fig \ref{fig:press_study} shows the enthalpy {and volume as a function of the pressure across the  lowest enthalpy structures from the GASP run at 50GPa and 150GPa. Four of the five lowest enthalpy structures from both the GASP runs had the same symmetry. The $P6_3mc$ structure shows almost no variation in enthalpy and volume with respect to the $F\Bar{4}3m$ structure because they are related through stacking faults. The Pt atoms in $F\Bar{4}3m$ form an ABC type closed pack stacking whereas they form an ABA type closed pack stacking in $P6_3mc$. The structure with $F\Bar{4}3m$ space group has the lowest enthalpy and thus the ambient condition ground-state structure remains the most stable structure up to 150 GPa.} 

The absence of pressure induced structural transitions is not surprising given that the ambient pressure structures of both Pt and Be individually exhibit a remarkable stability under pressure.  
Platinum remains in the ambient pressure fcc structure to at least \SI{304}{GPa}~\cite{mao_pressure_1990}.
The stability of fcc Pt under pressure can be attributed to the fact that its neighbor to the right in the periodic table, Au, also adopts the fcc structure.
Pressure is well known to induce a transfer of $s$-electrons into $d$-states in several elements.
This $s\rightarrow d$ transfer often leads to structural transitions.
As $s\rightarrow d$ transfer occurs in Pt, it becomes more Au-like.
However, Au is also in the fcc structure, so no structural transition occurs~\cite{ahuja_electronic_1994}.
Beryllium remains in the ambient pressure HCP phase to at least \SI{174}{GPa}~\cite{mcmahon_high-pressure_2006}.  In the heavier alkaline earth elements (Ca, Sr, and Ba), pressure-induced $s\rightarrow d$ transfer leads to the appearance of complex crystal structures at high pressure~\cite{hamlin_superconductivity_2015}.  However, Be has no nearby $d$ states.  Lithium, adjacent to Be, also has no nearby $d$ states, yet exhibits a number of complex crystal structures under relatively low pressures, due to the overlap of the $1s^2$ core electrons.  However, the $1s^2$ core of Be (\SI{1.38}{pm}) is substantially smaller than that of Li (\SI{1.86}{pm})~\cite{waber_orbital_1965}, and hence core overlap will not begin to occur until much higher pressures than for Li.

{ The arguments above involve solids formed from the isolated elements Be and Pt.
Of course, in a compound, 
there could always be another structure that exhibits a smaller volume of lower energy, hence decreasing the enthalpy. However, the PtBe$_5$ compound is a comparably close packed. Therefore, a priori, we did not anticipate a phase transformation and confirmed this expectation by a genetic algorithm search.
A concrete example of such a mechanism is the structural transition in Hume-Rothery  phases~\cite{mott_theory_1958}, driven by Fermi-sphere Brillouin-zone interactions,  as well as pressure induced structural transitions~\cite{Degtyareva_2006}.
The basic idea is that when the Fermi sphere is close to the Brillouin zone boundary, a gap can open, thus lowering the energy.
A structural transition can then occur if the transition brings additional portions of the Brillouin zone into close proximity with the Fermi sphere (thus further lowering the energy).
Due to the low carrier density of \ch{Be5Pt}, this mechanism is unlikely to be relevant, however.
}

\section{ Pressure dependence of resistivity }
\label{sec:pressure}
\begin{figure}
    \centering
    \includegraphics[width=\columnwidth]{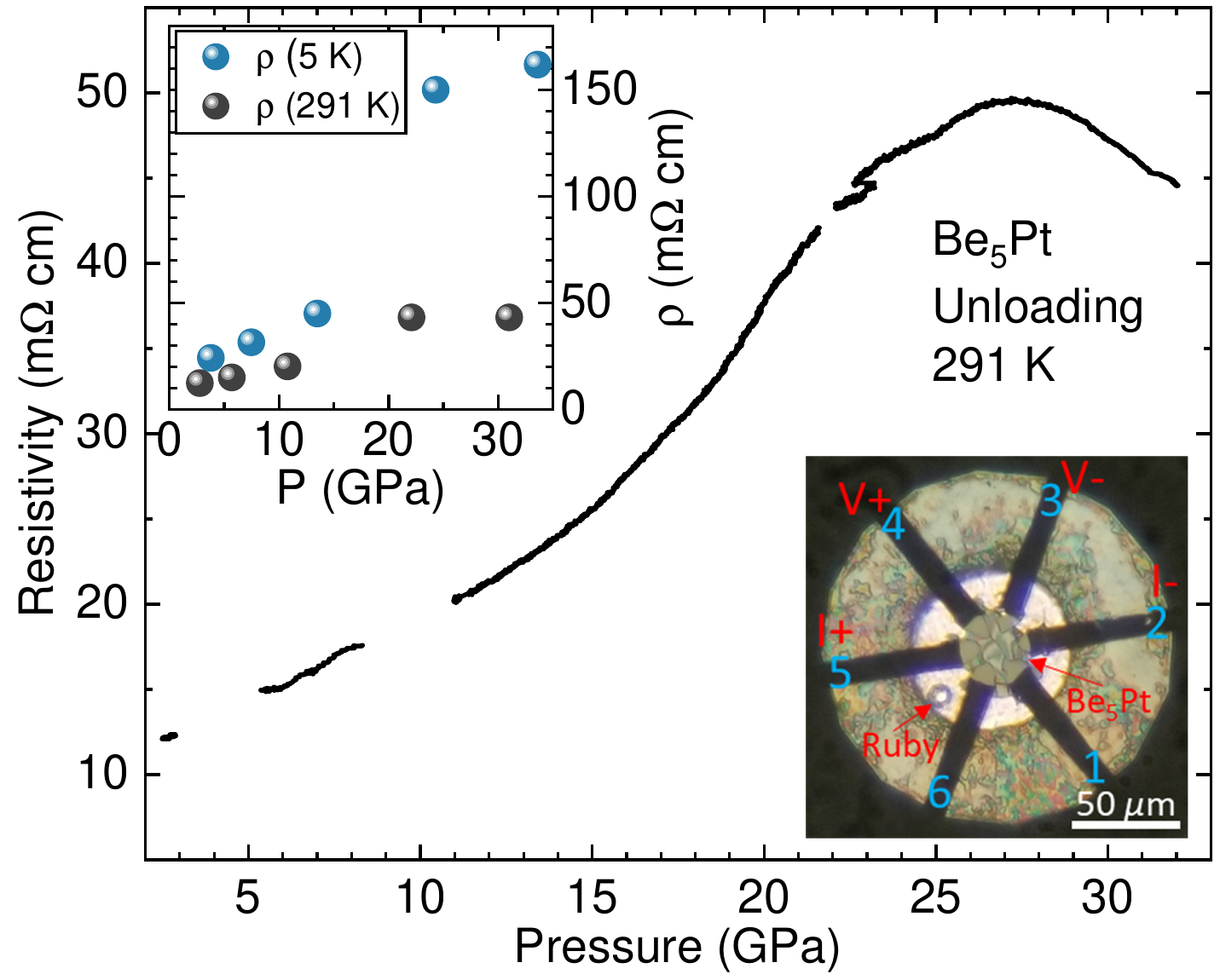}
    \caption{Resistivity of \ch{Be5Pt} versus pressure from $\sim$ 32 down to 2 GPa (unloading) at room temperature.  Inset represents the resistivity values at five different pressures for 5 and 291 K. Notice that pressure change due to change in temperature (291 down to 5 K) is relatively small, for example $\sim$ 1 GPa at 2.8 GPa and $\sim$ 2.6 GPa at 31 GPa, indicating the stability of the pressure cell. Bottom right inset shows the photograph of \ch{Be5Pt} sample ($\sim$ 40 $\times$ 40 $\times$ $\SI{5}{\micro\m^3}$) along with a ruby for pressure calibration, a soapstone insulation (a bright area surrounding the sample), a 316 stainless steel metal gasket with a hole ($\sim$ $\SI{80}{\micro\m}$ in diameter), and six leads (tungsten) configuration. Leads 3 and 4 were used to measure voltage drop, whereas leads 2 and 5 were used to provide current flow.}
    \label{fig:PtBe5_rhoP_V1}
\end{figure}

\begin{figure}
    \centering
    \includegraphics[width=\columnwidth]{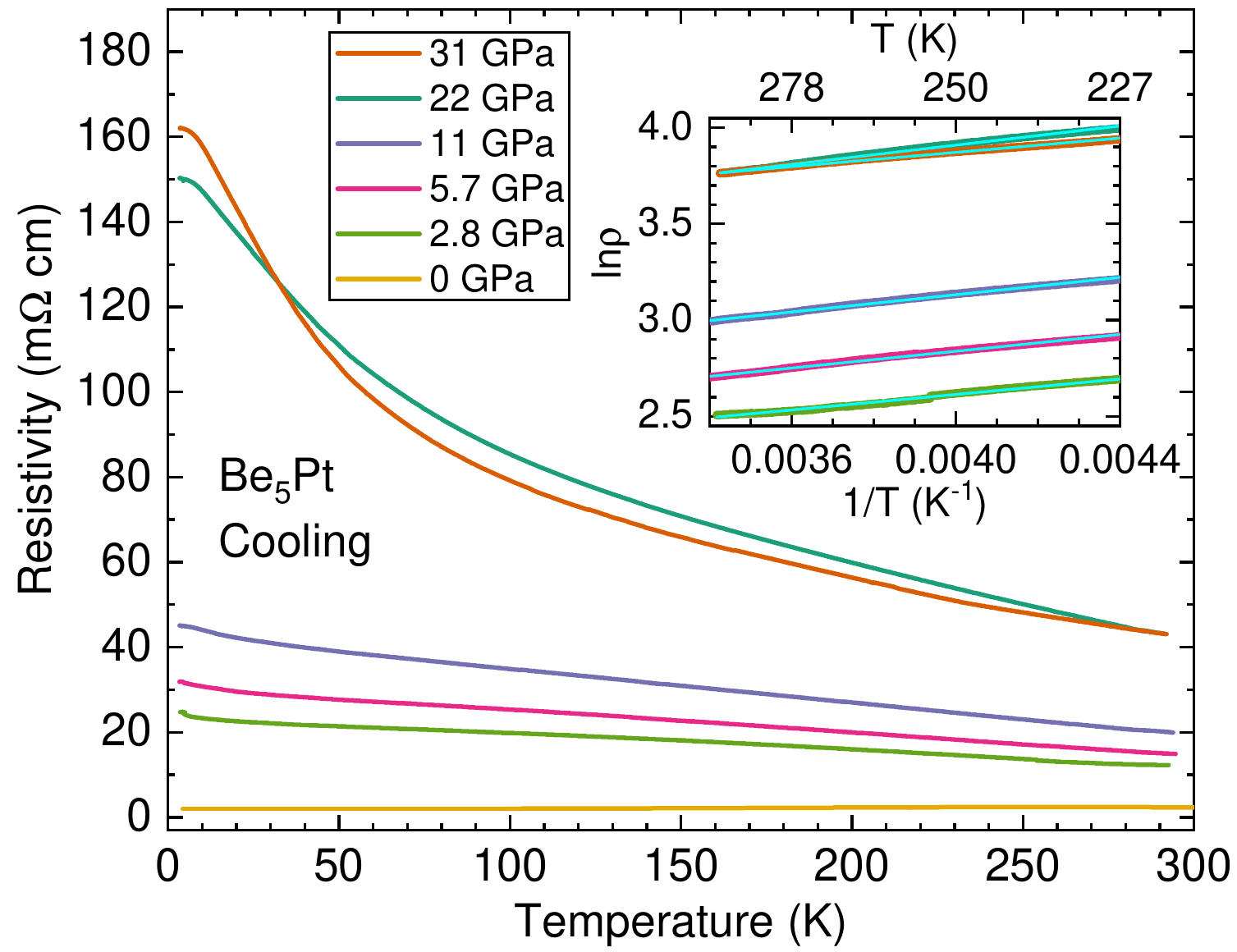}
    \caption{Resistivity versus temperature curves (cooling) for \ch{Be5Pt} at 31, 22, 11, 5.7, 2.8, and 0 GPa (measured at room temperature). Data at 0 GPa is adopted from Fig.~\ref{fig:transport_expt}. Inset shows  $\ln{\rho}$ versus 1/$T$ plot to calculate the band gap energy ($E$\textsubscript{g}) using the Arrhenius relation, $\ln{\rho(T)}$ = $\ln{\rho_0}$ + $E\textsubscript{g}/2k\textsubscript{B}T$. Cyan  lines are the linear fits to the data from 0.0034 K\textsuperscript{-1} (294 K) to 0.0044 K\textsuperscript{-1} (227 K), where each slope gives the corresponding $E\textsubscript{g}/2k\textsubscript{B}$ value.}
    \label{fig:PtBe5_rhoT_V1}
\end{figure}

For the high-pressure resistivity measurements, a micro-sized \ch{Be5Pt} sample ($\sim$ 40 $\times$ 40 $\times$ $\SI{5}{\micro\m^3}$) was cut from a larger piece of polycrystalline sample and placed in a gas-membrane-driven diamond anvil cell (OmniDAC from Almax-EasyLab) along with a ruby ($\sim$ $\SI{10}{\micro\m}$ in diameter) for pressure calibration~\cite{chijioke_ruby_2005}. Two opposing diamond anvils (0.15 and 0.5 mm central flats) were used, one of which was a designer-diamond anvil (0.15 mm central flat) with six symmetrically deposited tungsten microprobes in the encapsulated high-quality-homoepitaxial diamond~\cite{weir_epitaxial_2000}. A 316 stainless steel metal gasket was pre-indented from $\sim$ 150 to $\SI{25}{\micro\m}$ in thickness with a hole ($\sim$ $\SI{80}{\micro\m}$ in diameter), which was filled with soapstone (steatite) for insulating the sample and which also served as the pressure-transmitting medium. For the resistivity calculation, we used the van der Pauw method, (assuming an isotropic sample in the measurement plane), ${\rho}$ = ${\pi}$tR/$\ln{2}$, where t is the sample thickness ($\sim$ $\SI{5}{\micro\m}$) with a current of 1 mA. The high-pressure resistivity cell was placed inside a customized continuous-flow cryostat (Oxford Instruments). A home-built optical system attached to the bottom of the cryostat was used for the visual observation of the sample and for the measurement of the ruby manometer. The pressure was applied from $\sim$ 1 (initial loading) to 32 GPa slowly for $\sim$ 5 hours for the pressure-dependent resistivity measurement at room temperature and then released to each pressure (Fig.~\ref{fig:PtBe5_rhoP_V1}) for the temperature-dependent resistivity measurements in the range from $\sim$ 295 down to 1.8 K (Fig.~\ref{fig:PtBe5_rhoT_V1}).

Figure~\ref{fig:PtBe5_rhoP_V1} shows room-temperature resistivity data under high pressure from $\sim$ 32 down to 2 GPa during decompression. The resistivity data during compression is not shown because lead 1 (see Fig.~\ref{fig:PtBe5_rhoP_V1} inset), used for one of the current leads, had a short with the metal gasket, which might include additional resistance in the data other than the sample. The leads for unloading pressure are without any short. It is known that resistivity obtained from decompression provides more accuracy than from compression because sample thickness is relatively well-kept during decompression~\cite{seagle_iron_2013}. The resistivity of \ch{Be5Pt} at room temperature increases  under high pressures, which indicates the band gap energy ($E$\textsubscript{g}) also increases according to the Arrhenius relation, ${\rho(T)}$ = ${\rho_0}$ $\exp{(E\textsubscript{g}/2k\textsubscript{B}T)}$ at a fixed $T$. This is consistent with the band structure calculations for greater than ambient pressure in Fig.~\ref{fig:esp}, showing the band gap opens further with increasing pressures to $\sim$ 27 GPa.  The  inset in Fig.~\ref{fig:PtBe5_rhoP_V1} compares the resistivity at 5 and 291 K for selected pressures. The anomaly $\sim$ 23 GPa in Fig.~\ref{fig:PtBe5_rhoP_V1}, which shows pressure reversing, is from the adjustment of pressure determination of the ruby manometer, whereas the discontinuities in the resistivity curve are due to the changes in pressures before and after temperature-dependent resistivity measurements.

The temperature-dependent resistivity curves of \ch{Be5Pt} are shown in Fig.~\ref{fig:PtBe5_rhoT_V1} for five unloading pressures at 31, 22, 11, 5.7, and 2.8 GPa, including a separate ambient pressure measurement (see Fig.~\ref{fig:transport_expt}). Remarkably, the resistivity curve at ambient pressure is nearly flat over a wide temperature range compared with the high pressure resistivity curves, as discussed above. The resistivity at 2.8 GPa begins to increase monotonically with decreasing temperature, which corresponds to a typical semiconducting behavior. With further increasing pressure, the increase in temperature-dependent resistivity gets larger and larger  (see Fig.~\ref{fig:PtBe5_rhoP_V1} inset). This enhancement in the semiconducting behavior under high pressures to 31 GPa is again in a good agreement with the DFT calculations of pressure effects on \ch{Be5Pt} in Sec. III.  The fits of the Arrhenius equation to the high pressure resistivity curves (ln${\rho}$ versus 1/$T$ plot) are shown in the Fig.~\ref{fig:PtBe5_rhoT_V1} inset in the temperature range from $\sim$ 227 to 294 K to calculate the band gap energy ($E\textsubscript{g}$). The estimated $E\textsubscript{g}$ values are $\sim$ 35, 38, 39, 43, and 32 meV at 2.8, 5.7, 11, 22, and 31 GPa, respectively, which shows a good  qualitative agreement with the calculated true gaps shown in Fig.~\ref{fig:esp}.

\section{Conclusions}
Through a series of measurements and theoretical calculations, we have argued that the little-studied intermetallic semiconductor \PB5 is in fact an extremely unusual member of its materials class. While not superconducting like its cousin Be$_{21}$Pt$_5$, our electronic structure calculations suggest that at ambient pressure it is semiconducting with one of the smallest gaps of any intermetallic, an indirect gap of order 3 meV according  to DFT calculations we presented here.  Insulating temperature  dependence is {\it not} observed in low-temperature transport, however; rather the resistivity is flat at ambient pressure over a range of roughly 100K, together with a strongly $T$-dependent Hall effect.

The calculated band structure near the Fermi level has  hole and electron bands resembling Dirac cones whose extrema are accidentally nearly degenerate.  In addition, there is a Weyl loop structure quite close to the Fermi level, such that a small amount of doping might allow observation  of topological quantization.  Here we studied the transport  in the quasi-Dirac cone bands and developed a model to explain the $T$-independent resistivity  together with the sign and unusual $T$ dependence of the observed Hall coefficient.  Agreement with experiment requires assumptions about the influence of the Be inclusions found in current samples, however.    

Calculations were performed under pressure using a genetic algorithm    structural relaxation method, which failed to find any phase transitions from the ambient pressure $F{\bar 4}3m$ structure up to 150 GPa, consistent with resistivity  measurements performed in a diamond anvil cell up to this pressure.  However, the temperature dependence of the resistivity evolved significantly,  consistent with the opening of the gap found in DFT calculations.  

In summary, we have performed a close investigation of the properties of the intermetallic semiconductor \PB5, and shown that it is a remarkable member of this class, primarily  because of an extraordinarily small accidental gap, surrounded by a somewhat larger pseudogap feature in the density of states.  Additional unusual features of the band structure include Weyl loops close to the Fermi level; we have suggested that doping the compound with a small amount of gold may allow one to bring this feature to the Fermi level. 
~ \\
\vskip 0.2cm 
~ 
{\bf Acknowledgements.}  The authors are grateful for useful discussions with Y. Grin and Young-Joon Song.  Work on this project was supported by the US Department of Energy Basic Energy Sciences under Contract
No. DE-SC-0020385. L. F. acknowledges support by the European Union’s Horizon 2020 research and innovation programme through the Marie Sk\l{}odowska-Curie grant SuperCoop (Grant No 838526).

\appendix
\section{Transport} \label{Transport}

We consider a two-band toy model with two Dirac-like bands. For the moment, we neglect the tiny band gap present in the \PB5 system, i.e. we fix $\Delta_e =\Delta_h = 0$, and further assume that $\mu=0$ at zero temperature. The band dispersion is then $\epsilon_\alpha =\pm \gamma_{\alpha}  k$, where $\alpha= e,h$ and ${\bf k}$ is measured with respect to the touching point of the band. This will yield a linear density of states in energy $N_\alpha(\epsilon)=|\epsilon|/(\gamma_\alpha^2 \pi)$. 
To account for the particle-hole asymmetry of the density of states found in the band structure calculations for \PB5 we need $\gamma_h >\gamma_e$. 
Note that because the Dirac cone is now asymmetric, the chemical potential $\mu$ depends significantly on $T$ and must be calculated self-consistently.

Assuming that the transport at low temperatures is dominated by a small number of weakly scattering pointlike defects with concentration $n_i$ and potential $u$ the scattering rate reads~\cite{Ando98}
  \begin{eqnarray}\label{Eq:selfenergy}
    {1\over \tau_\alpha(\omega)}&=& 2\Sigma''(\omega) = 2n_i u^2 \sum_{\bf k} {\rm Im~} G({\bf k},\omega) \nonumber\\
    &=&  \pi n_i u^2 \int d\epsilon \ N_\alpha(\epsilon) \delta (\omega +\epsilon) = \Gamma^0_\alpha |\omega|,
  \end{eqnarray}
where we introduced the dimensionless parameter $\Gamma^0_\alpha = n_i u^2/\gamma_\alpha^2$, characterizing the scattering strength.
In this and subsequent expressions we have set both the lattice parameter $a=1$ and the reduced Planck constant $\hbar=1$.

The longitudinal conductivity is given by the sum of the electron and hole contribution $\sigma_{xx}= \sum_{\alpha} \sigma^\alpha_{xx}$ with

\begin{eqnarray} 
\sigma^{\alpha}_{xx} &=& \frac{e^2 v_\alpha^2}{2} \int \frac{d^2{\bf k}}{(2\pi)^2}  \  \tau_\alpha (\epsilon_{\bf k})  \  \bigg(-\frac{\partial f}{\partial \epsilon_{\bf k}}\bigg)\nonumber\\ &=&
\frac{e^2 v_\alpha^2}{2} \int d\epsilon \ N_\alpha(\epsilon) \ \tau_\alpha(\epsilon)  \  \bigg(-\frac{\partial f}{\partial \epsilon}\bigg)\nonumber\\ &=&
\frac{e^2 \lambda_\alpha }{2  \pi \Gamma^0_\alpha},
\label{app:sigma_xx_v2}
\end{eqnarray}
where $v_\alpha = \pm \gamma_{\alpha}$ is the velocity for the $\alpha= e,h$ band and $\lambda_\alpha$ a numerical factor coming from the evaluation of the integral and that in our case is $\lambda_h > \lambda_e$. 
We recover the result quoted in the main text that shows that the $T$-dependence of the longitudinal conductivity vanishes to leading order. 
The total conductivity is dominated by the contribution of the holes that is in our model $\sim 9$ times larger than that of the electrons. This effect results from the strong hole-electron anisotropy of the bands that forced us to assume $\gamma_h >\gamma_e$ and pushes the chemical potential to negative value as the temperature increases.

\begin{figure}[tbh]
  \includegraphics[width=\columnwidth]{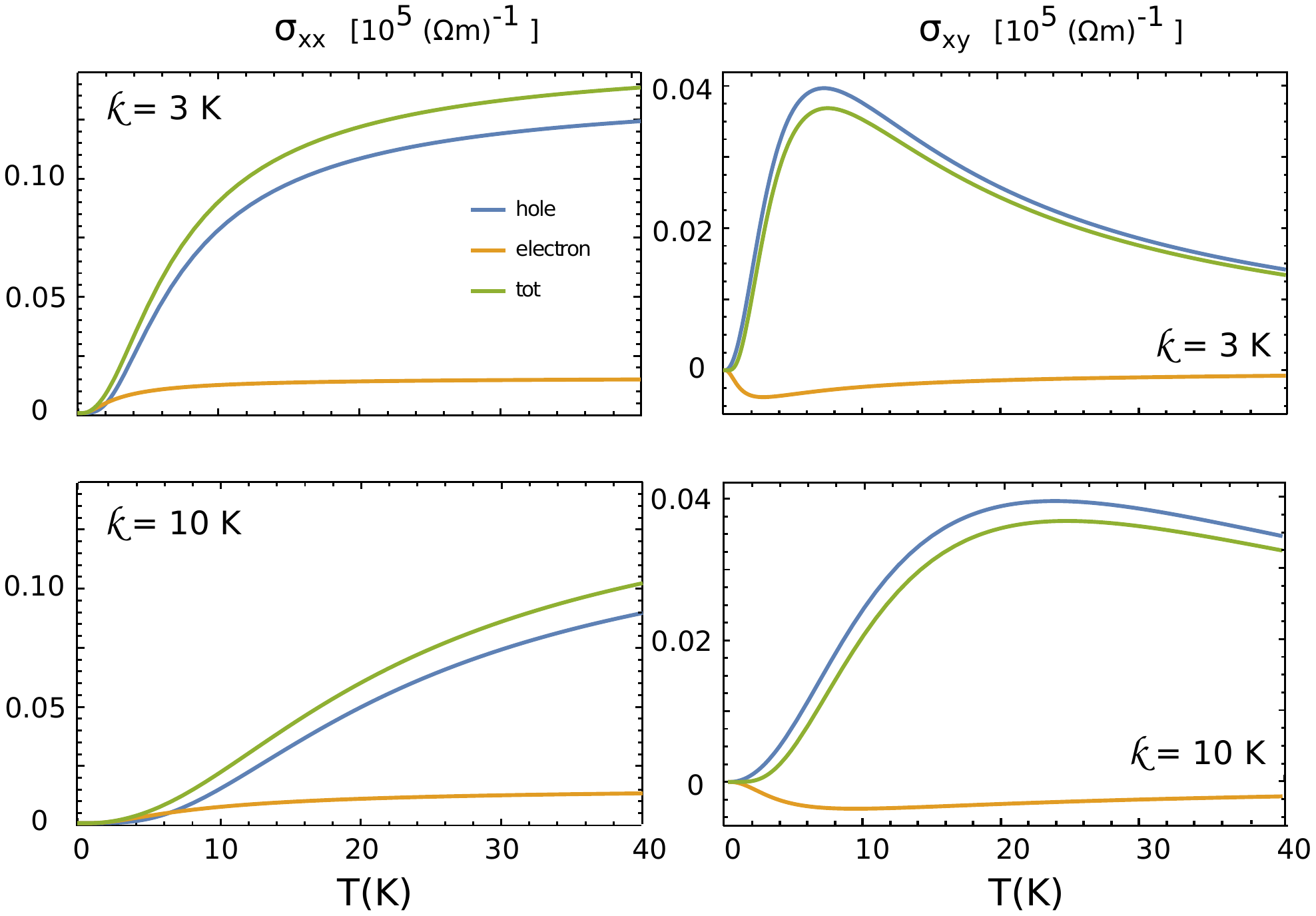}
  \caption{Temperature dependence of the hole and electron band contributions to $\sigma_{xx}(H)$ and $\sigma_{xy}$ for $\kappa_e=3, \ 10$ K. To estimate the temperature independent term $e^2/(2 \pi \Gamma_\alpha^0)$, we fix  $\Gamma^0_\alpha$ such that the total longitudinal conductivity at at zero field, Eq.(\ref{app:sigma_xx_v2}), fits the experimental value $\sigma_{xx} = 0.5 \ 10^5 (\Omega \text{m})^{-1}$. The anisotropy of band structure, encoded in the parameters $\gamma^2_h \sim 4 \gamma^2_e$, and in the temperature shift of the chemical potential to negative values, makes the transport dominated by the hole band over the whole range of temperature. $\kappa^2_\alpha$ acts as a cut-off for the divergence of the $\sigma^{\alpha}_{xy}$}
    \label{fig:Cond}
\end{figure}
  
The Hall coefficient in a two band system is given by 
\begin{equation}
    R_H={\sigma_{xy}\over H \ (\sigma_{xx}^2 + \sigma_{xy}^2)},
    \label{app:RH}
\end{equation}
where $H$ is the magnetic field,  $\sigma_{xx} = \sum_\alpha\sigma^\alpha_{xx}$ is the total longitudinal conductivity and $\sigma_{xy} = \sum_\alpha\sigma^\alpha_{xy}$ the transverse one. 
Following the same procedure as in Eq.(\ref{app:sigma_xx_v2}), the transverse conductivity for each band is given by\cite{Ando2006}
\begin{eqnarray}
\sigma^\alpha_{xy} &=&  - \frac{e^2 v_\alpha^2}{2}\int  d\epsilon \ N_{\alpha}(\epsilon) \ \frac{{\omega^\alpha_c}(\epsilon) \  \tau_\alpha^2 (\epsilon)}{1+({\omega_c}^\alpha(\epsilon) \  \tau_\alpha (\epsilon))^2} \  \bigg(-\frac{\partial f}{\partial \epsilon}\bigg) \nonumber \\
&=&  \frac{e^2}{2 \pi \Gamma^0_\alpha} \ \kappa_\alpha^2 \int_{-\infty}^{0} d\epsilon \ \frac{\epsilon^2}{ \epsilon^4  + \kappa_\alpha^4 } \  \bigg(-\frac{\partial f}{\partial \epsilon }\bigg).
\label{app:sigmaxy}
\end{eqnarray}
Here $\omega^\alpha_c = e H v_\alpha^2/\epsilon$ is the energy dependent cyclotron frequency. It is negative for the hole band, $\epsilon_h<0$, and positive for the electron one, $\epsilon_e>0$. We further introduced the parameter $\kappa_\alpha$ such that $\omega^\alpha_c(\epsilon)\tau_\alpha(\epsilon) = \kappa_\alpha^2 / \epsilon^2$, i.e. 
\begin{equation}\label{app:kappa}
\kappa_\alpha^2 =\frac{e H v_\alpha^2}{\Gamma^0_\alpha}.
\end{equation}
Since $1/\tau_\alpha(\epsilon)$ and $N_\alpha(\epsilon)$ vary as $\epsilon$, the integral without the $({\omega^\alpha_c} \tau_\alpha)^2$ term in the denominator is actually divergent as $1/\epsilon^2$.  Thus this term is required, unlike the ordinary Hall effect for metallic systems, even in the low field limit, and determines the $\sigma^\alpha_{xy}\sim T^2 $ dependence as $T\rightarrow 0$.  To determine the crossover between low and high-temperature behavior, we must estimate the value of $\kappa_\alpha$ from Eq.(\ref{app:kappa}).
Notice that the T-behavior of the two bands change over different energy scales as one can see in Fig.(\ref{fig:Cond}) where we show the temperature-dependence of the transverse conductivity together with the electron and hole contributions for two representative values of $\kappa_e$. Here we used $\gamma_h \sim 4 \gamma_e$, as extracted from the density of states in Fig.(\ref{fig:es}) approximated as linear as shown in the schematic of Fig.(\ref{fig:schematic_bands}). As a consequence we have that $\kappa_h \sim 4 \kappa_e$, thus the low-high temperature crossover appears at lower temperature for the electron term with respect to the hole term. The sign of the transvers conductivity is positive due to the dominant contribution of the hole band.

\begin{figure}[tbh]
  \includegraphics[width=0.85\columnwidth]{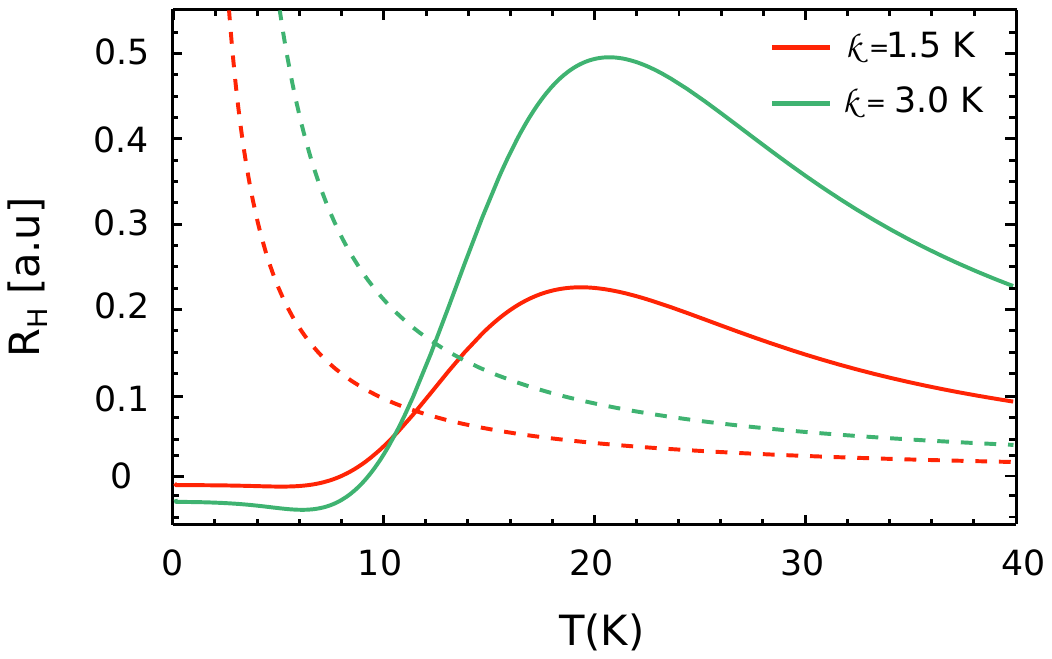}
  \caption{Temperature dependence of $R_H$ for the two-band Dirac model. Dotted lines: result obtained using $\Delta_e=\Delta_h=0$, $\mu(T=0)=0$ for $\kappa_e=1.5, 3.0$ K. We find that the transport is dominated by the hole band and we get a divergent $R_H(T)$ for $T\rightarrow 0$. Continuous lines: result obtained using $\Delta_e=\Delta_h= 0.5$ meV and $\mu(T=0)=3.4$ meV. If the chemical potential at $T=0$ lies inside the electron band the divergence at low-T is suppressed by the contribution of the electron band and we get an $R_H$ that  increases as T is increased over the 3 - 20 K range.}
  \label{fig:Hall}
\end{figure}

Since in the evaluation of $\sigma_{xy}$  we account for the full magnetic field dependence, i.e. we retain the full denominator in Eq.(\ref{app:sigmaxy}), when computing the Hall coefficient via Eq.(\ref{app:RH}), we need to use the same approximation for the field-dependent longitudinal conductivity $\sigma_{xx}$, i.e. 
\begin{eqnarray}
\sigma^\alpha_{xx} (H)&=&  - \frac{e^2 v_\alpha^2}{2}\int  d\epsilon \ N_{\alpha}(\epsilon) \ \frac{ \tau(\epsilon)}{1+{\omega_c}^2(\epsilon) \  \tau^2 (\epsilon)} \  \bigg(-\frac{\partial f}{\partial \epsilon}\bigg) \nonumber \\
&=&  \frac{e^2}{2 \pi \Gamma^0_\alpha}  \int_{-\infty}^{0} d\epsilon \ \frac{\epsilon^4}{ \epsilon^4  + \kappa_\alpha^4 } \  \bigg(-\frac{\partial f}{\partial \epsilon }\bigg)
\label{app:sigmaxx_h}
\end{eqnarray}
whose temperature-dependence is shown in Fig.(\ref{fig:Cond} for the same values of $\kappa_\alpha$ used in the $\sigma_{xy}$ calculation. It is evident that the total contribution of both the longitudinal and transverse conductivities come essentially form the hole band regardless of the value of $\kappa_\alpha$. The strong anisotropy of the Dirac cone considered here reduces the two-band model to a single hole-like band model. As a consequence, the Hall coefficient reduces to $R_H \sim \sigma_{xy}^h/(\sigma_{xy}^h)^2 +(\sigma_{xx}^h)^2$, that regardless of the exact value of the $\kappa_\alpha$ parameter used, present a divergent behavior as $T\rightarrow 0$ is approached, see dotted lines in Fig.(\ref{fig:Hall}). The anomalous temperature dependence of $\sigma_{xx}(H)\sim T^4$ at low $T$ is at the root of this behavior in the current model.

The inclusion of a small gap in the model, does not change qualitatively the low-temperature behavior of the conductivities and of the Hall coefficient. This new energy scale could change the relative weight of the bands especially if $\kappa_\alpha$ is of the same order of the gap. However, this effect is small with respect the suppression of the electron band contribution due to the strong temperature dependent chemical potential shift to negative values that makes the physics still dominated by the hole-like band over the whole temperature range. 

An effective parameter that crucially affects the temperature dependence of the Hall coefficient is, instead, the doping. In fact, if we assume $\mu(T=0) > 0$, such that the Fermi level at zero temperature lays at the edge of the electron band, or further inside the band, we recover a not negligible contribution of the electron band at very low T that effectively cuts off the divergence of the the Hall coefficient at low temperature, see Fig.(\ref{fig:Hall}) where we show the results for a specific case with $\Delta_e=\Delta_h= 0.5$ meV and $\mu(T=0)=3.4$ meV for two different values of $\kappa_{e}$. As a consequence the Hall coefficient is now increasing as T is increased up to $\sim 20$ K as seen in experiment.
Notice that the crossover between the divergent $R_H(T)$ of the un-doped model to the $R_H(T)$ increasing with temperature of the doped case is controlled by the position of the Fermi level at zero temperature with respect the electron band, thus it is also affected by the gap values. Moreover the values of $\kappa_\alpha$ used in the calculation influences the range of temperature over which $R_H$ increases, i.e. the position of the peak as shown in Fig.(\ref{fig:Hall}). This result thus requires fine tuning, which is probably unreasonable given the other uncertainties in the analysis. In the main text we explicitly discuss the possible role of the existence of Be inclusions in the samples. 

\begin{figure}[!ht]
  \includegraphics[width=0.8\columnwidth]{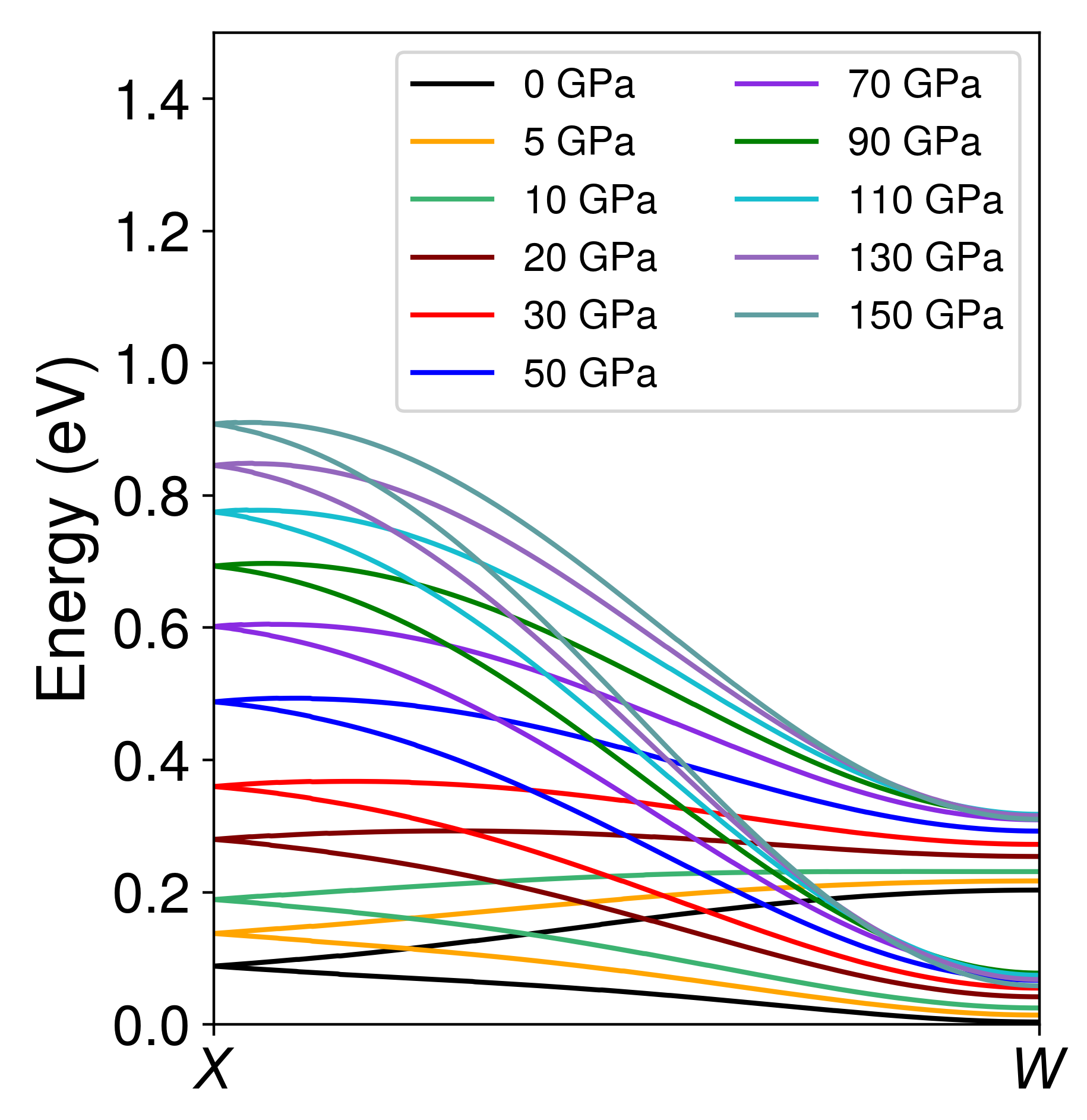}
  \caption{{ In the figure above, the band dispersions of the two lowest unoccupied bands under eleven different pressures are shown. The two fold degeneracy at $X$ ponit gives rise to the first DOS peak above the Fermi level.}}
  \label{peak}
\end{figure}

{ \section{Pressure dependence of the lowest unoccupied DOS peak} \label{dospeak}
We have shown in Fig. 7 (b) of the main text that the first peak of the unoccupied DOS is strongly pressure dependent. To understand the origin of the pressure dependence, we plot the band dispersion of the two lowest conduction bands along $X \rightarrow W$ in Fig.~\ref{peak}. The first DOS peak
above the Fermi level is the result of the two conduction bands (Pt 5d bands) touching at $X$. As pressure increases, the two bands touch at progressively higher energy. As a result, the DOS peak is pushed up and a significant amount of spectral weight is transferred away from the Fermi level.}

\section{Dependence of the bandgap on the exchange-correlation functional}
Because the true spectral (indirect) gap of Be$_5$Pt of 3 meV found in the calculations presented is unusually small, we have checked this result by calculating the same quantity using several exchange-correlation functionals, including LDA, PBE, the strongly constrained and appropriately-normed exchange-correlation functional (SCAN)~\cite{scan1,scan2}, as well as the hybrid functional HSE06~\cite{HSE06}. The predicted bandgaps shown in Tab.~\ref{tab:gaps} are consistently very small, ranging from 3 to 22 meV, except for the HSE06 result of 297 meV.  We note further that in Ref.~\cite{amon_interplay_2019}, a larger bandgap of 85~meV was estimated from the DOS calculated using the FPLO code. It is well known that while HSE06, with the conventional choice of Hartree-Fock (HF) mixing coefficient of $\alpha=1/4$, works well for molecular systems and simple semiconductors, it can significantly overestimate the bandgap in systems with large dielectric constants and small bandgaps~\cite{Gerosa2017}. In fact, from our Arrhenius analysis of the resistivity presented in Fig.~\ref{fig:PtBe5_rhoP_V1}, the larger HSE06 bandgap is inconsistent with the data, which suggest a gap of at most $\sim30$~meV. Reducing the amount of HF mixing brings the bandgap closer to the experimental value, indicating that Be$_5$Pt has a sizeable electric susceptibility. We note further that the shape of the bandstructure and the density of states does not change significantly among all these different functionals, such that the analysis presented here is expected to be robust against small errors in the gap value.  

\begin{table}[t]
  \caption{\label{tab:gaps} Calculated bandgaps in meV of Be$_5$Pt using several exchange-correlation functionals and the VASP and Wien2k codes. For the hybrid functional HSE06, we vary the amount of HF mixing from the default value of $\alpha=1/4$ to a smaller value of $1/8$.  All results include spin-orbit coupling interactions.}
  \begin{ruledtabular}
  \begin{tabular}{c|cccccc}
     & LDA & PBE & SCAN&HSE06 & HSE06 \\
     & & & & $\alpha=1/4$ & $\alpha=1/8$ \\
     \hline
     VASP & 22 & 8 & 12 & 297 & 160 \\
     WIEN2k & 13  & 3 & 4 &-- & --\\
  \end{tabular}
  \end{ruledtabular}
\end{table}

\bibliography{PtBe5}

\end{document}